\documentclass[journal=jacsat,manuscript=perspective,layout=twocolumn]{achemso}

\usepackage[version=3]{mhchem} % Formula subscripts using \ce{}

\usepackage{multirow}
\usepackage[acronym]{glossaries}
\usepackage{svg}
\makeglossaries
\usepackage{enumitem}
\usepackage{booktabs}
\usepackage{mathtools}
\newacronym{DFT}{DFT}{Density Functional Theory}
\newacronym{GO}{GO}{Geometry Optimization}
\newacronym{S2NS}{S2NS}{Structure to Next Structure}
\newacronym{S2RS}{S2RS}{Structure to Relaxed Structure}
\newacronym{S2EF}{S2EF}{Structure to Energy and Forces}
\newacronym{IS2RE}{IS2RE}{Initial Structure to Relaxed Energy}
\newacronym{IS2RS}{IS2RS}{Initial Structure to Relaxed Structure}
\newacronym{GNNs}{GNNs}{Graph Neural Networks}
\newacronym{OC20}{OC20}{Open Catalyst 2020 Dataset}
\newacronym{OCP}{OCP}{Open Catalyst Project's}
\newacronym{EwT}{EwT}{Energy within Threshold}
\newacronym{ADwT}{ADwT}{Average Distance within Threshold}
\newacronym{AFbT}{AFbT}{Average Force below Threshold}
\newacronym{DwT}{DwT}{Distance within Threshold}
\newacronym{FbT}{FbT}{Forces below Threshold}
\newacronym{EFwT}{EFwT}{Energy Forces within Threshold}
\newacronym{ML}{ML}{Machine Learning}
\newacronym{MLP}{MLP}{ML potentials}
\newcommand{\mr}[2]{\multirow{#1}{*}{#2}}
\newcommand{\mc}[3]{\multicolumn{#1}{#2}{#3}}
\newcommand{\angstrom}{\text{Å }}
\author{Adeesh Kolluru}
\affiliation{Co-first Author}
\alsoaffiliation{Department of Chemical Engineering, Carnegie Mellon University}
\author{Muhammed Shuaibi}
\affiliation{Co-first Author}
\alsoaffiliation{Department of Chemical Engineering, Carnegie Mellon University}
\author{Aini Palizhati}
\affiliation{Department of Chemical Engineering, Carnegie Mellon University}
\author{Nima Shoghi}
\affiliation{Fundamental AI Research at Meta AI}
\author{Abhishek Das}
\affiliation{Fundamental AI Research at Meta AI}
\author{Brandon Wood}
\affiliation{Fundamental AI Research at Meta AI}
\author{C. Lawrence Zitnick}
\affiliation{Fundamental AI Research at Meta AI}
\author{John R Kitchin}
\affiliation{Department of Chemical Engineering, Carnegie Mellon University}
\author{Zachary W Ulissi}
\affiliation{Department of Chemical Engineering, Carnegie Mellon University}
\email{zulissi@andrew.cmu.edu}

\title[An \textsf{achemso} demo]
  {Open Challenges in Developing Generalizable Large Scale Machine Learning Models for Catalyst Discovery}

\begin{document}

% \begin{tocentry}

% Some journals require a graphical entry for the Table of Contents.
% This should be laid out ``print ready'' so that the sizing of the

% Inside the \texttt{tocentry} environment, the font used is Helvetica
% 8\,pt, as required by \emph{Journal of the American Chemical

% The surrounding frame is 9\,cm by 3.5\,cm, which is the maximum
% permitted for  \emph{Journal of the American Chemical Society}
% graphical table of content entries. The box will not resize if the
% content is too big: instead it will overflow the edge of the box.

% This box and the associated title will always be printed on a
% separate page at the end of the document.

% \end{tocentry}

\begin{abstract}
The development of machine learned potentials for catalyst discovery has predominantly been focused on very specific chemistries and material compositions. While effective in interpolating between available materials, these approaches struggle to generalize across chemical space. The recent curation of large-scale catalyst datasets has offered the opportunity to build a universal machine learning potential, spanning chemical and composition space. If accomplished, said potential could accelerate the catalyst discovery process across a variety of applications (\ce{CO2} reduction, \ce{NH3} production, etc.) without additional specialized training efforts that are currently required. The release of the \gls{OC20}\cite{chanussot2021open} has begun just that, pushing the heterogeneous catalysis and machine learning communities towards building more accurate and robust models. In this perspective, we discuss some of the challenges and findings of recent developments on \gls{OC20}. We examine the performance of current models across different materials and adsorbates to identify notably underperforming subsets. We then discuss some of the modeling efforts surrounding energy-conservation, approaches to finding and evaluating the local minima, and augmentation of off-equilibrium data. To complement the community's ongoing developments, we end with an outlook to some of the important challenges that have yet to be thoroughly explored for large-scale catalyst discovery.

% Recent developments in \gls{ML} models and the curation of large-scale catalyst
% datasets have pushed the field of heterogeneous computational catalysis towards
% more accurate \gls{MLP}. \gls{ML} models have evolved from being developed for
% specific chemistries and element types to large-scale generalized datasets.
% A universal \gls{MLP} has the potential to accelerate the catalyst discovery
% process across various applications (e.g. \ce{CO2} reduction, \ce{NH3} production, etc.)
% without additional specialized training efforts. In this perspective, we discuss
% the challenges associated with training multi-million parameter \gls{GNNs} on the
% recent \gls{OC20}, consisting of 200M+ adsorbate-catalyst \gls{DFT} calculations
% spanning 55 elements and 82 adsorbates. We begin with an overview of the progress
% on \gls{OC20} since its release. We then emphasize the challenges with training
% on various material types and adsorbates together. Lastly, we discuss the
% challenges in constructing tasks and metrics for the prediction of energy,
% forces and positions.
\end{abstract}

% progress and intro look interchangeable
\section{Introduction}
\begin{figure*}[t]
    \centering
    \includegraphics[width=\textwidth]{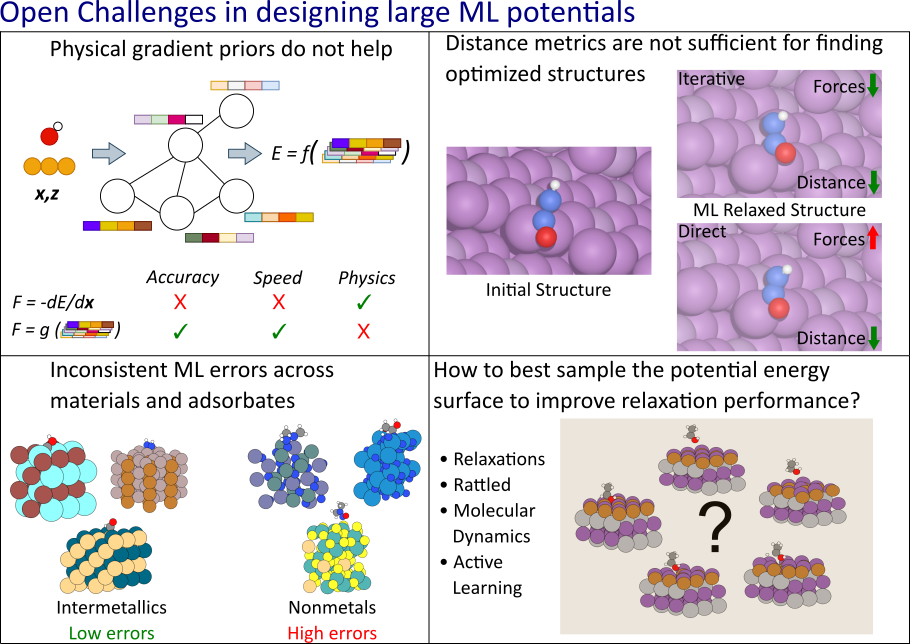}
    \caption{Summary of challenges associated with training on large dataset with large ML potentials discussed in the paper. \textit{Top left} Trade offs in direct and gradient GNN force predictions. \textit{Top right} An example system for a case where the distance metrics are relatively good for the direct approach but the force metrics are worse. \textit{Bottom left} Demonstration of inconsistent error across a metallic surface and a non-metal through an example. \textit{Bottom right} Augmenting existing relaxation datasets with off-equlibrium data can aid in relaxation performance.}
    % \vspace{-4mm}
    \label{fig:summary}
\end{figure*}
Catalysts have played a key role in the synthesis of everyday chemicals and fuels necessary for a 21st century society. As renewable energy prices continue to decrease, traditional chemical synthesis processes are being revisited for more sustainable alternatives. At the center of this, catalyst discovery plays a key role in the advancement of renewable energy processes and sustainable chemical production, i.e. ammonia for fertilizer and hydrogen production. Unfortunately, the search space for catalyst materials is enormous for even high-throughput experiments \cite{ren2018accelerated}. This presents a need for computational tools to simulate systems through quantum mechanical (QM) models like \gls{DFT}. QM approaches have made notable advancements in bridging computational results to experimental findings \cite{jose2006structure, rahali2021adsorption, xu2019dissolution, ktari2015design, rode2018synthesis, pause2001radical}. While effective, QM tools scale very poorly, $O(N^3)$ or worse in the number of electrons. The computational cost associated with QM tools render them infeasible to the scale of the systems and search space desired for catalyst discovery. As a result, the catalysis community has moved towards a more data driven approach \cite{agrawal2016perspective, guan2022machine, rosen2022realizing, himanen2019data, xu2021perspective}. With the QM data available, researchers are often interested in building machine learning surrogates for a particular chemical property \cite{tran2018active, back2019toward, ying2021unravelling, ge2020predicted}. Such efforts, however, were limited to the finite data available, often for a very specific chemistry or system, limiting the generalizability ability of such models \cite{guan2022machine, toniato2022grand}. Fortunately, as the community continues to curate larger, and more diverse datasets, machine learning models will continue to improve as they move towards larger, and more sophisticated architectures.

In the field of small molecules, a vast collection of datasets have been developed for varying use cases, including molecular dynamics simulations (MD17\cite{chmiela2017machine}, ANI-1\cite{smith2017ani}, COLL\cite{klicpera2020fast}) and quantum mechanical properties (QM9\cite{ramakrishnan2014quantum} Alchemy\cite{chen2019alchemy}). These datasets are often limited to a few (5-10) unique elements, on average 10-20 atoms per system, and training set sizes in the range of 10k-1M samples. In the field of heterogeneous catalysis, datasets are often much more limited with training set sizes between 100 - 50k \cite{andersen2019beyond, abild2007scaling, ma2017orbitalwise, noh2018active}. These datasets were often created for very specific applications involving a handful of small adsorbates (i.e. hydrogen containing adsorbates on transition metal surfaces, \ce{CO2} reduction catalysts, etc.). The release of \gls{OC20} marks a push towards a large, sparse collection of the material space. \gls{OC20} spans 55 unique elements, 82 adsorbates and includes a collection of unary, binary and ternary materials. A total of 1.28 million DFT relaxations were performed, comprising $\sim$260M single point evaluations of system energy and per-atom forces.

\gls{OC20} presented several practical tasks for the community to work towards. The most general of the tasks, \gls{S2EF} evaluates a model's ability to serve as a surrogate to DFT - predicting a configuration's energy and per-atom forces. \gls{IS2RE} asks to predict the relaxed state energy, given only the initial structure. \gls{IS2RS} explores how well the relaxed structure can be predicted given only the initial configuration. In the scope of \gls{OC20}, all energies were referenced to represent adsorption energy. For more details, we refer readers to the original manuscript \cite{chanussot2021open}.

In this perspective we shed light on the challenges of training Graph Neural Networks (GNNs) on large-scale datasets spanning material and composition space, illustrated in Figure \ref{fig:summary}. We begin with a quick overview on the current state of the community's progress and share some takeaways from what we have observed. We then discuss some telling trends on the performance of models across different adsorbates and material types. We discuss how different approaches and modeling decisions impact the prediction tasks and highlight the challenges associated with each.  Further, we explain what the accuracies in various proposed metrics mean and some of the challenges in analyzing them. Finally, we share our outlook on the direction the community is headed and what still remains to achieve a large scale, generalizable potential for catalyst discovery.
\section{Community progress in developing ML models for catalysis}
\begin{figure*}
    \centering
    \includegraphics[width=\textwidth]{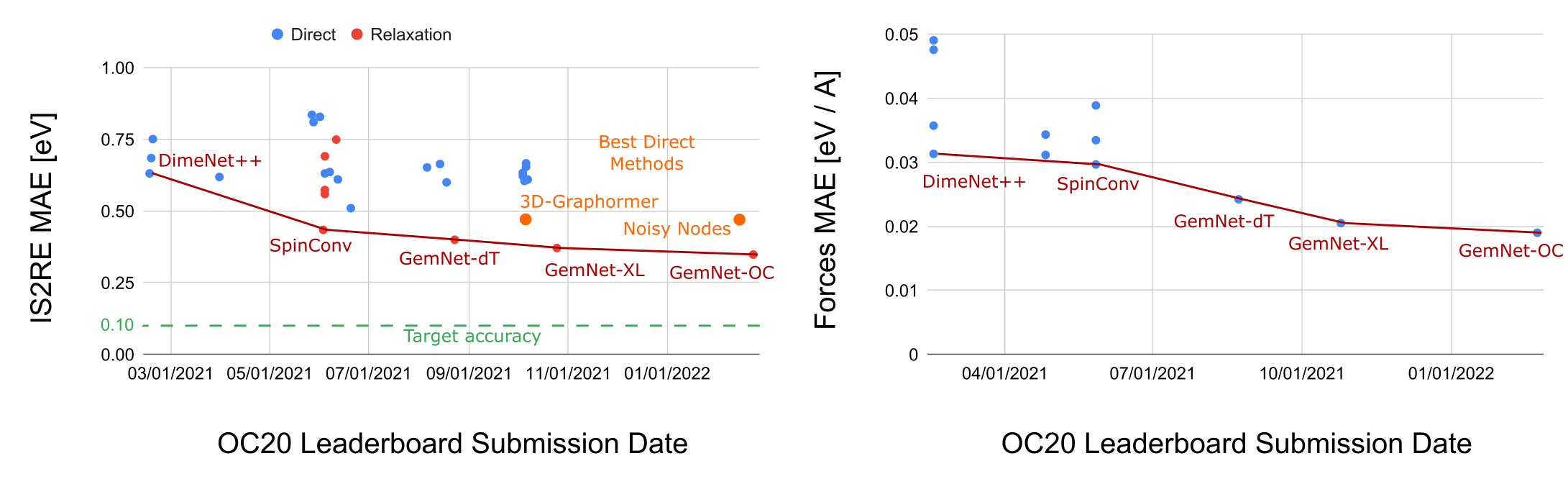}
    \caption{Community progress on the OC20 dataset since release. Left: \gls{IS2RE} performance for both direct and relaxation based approaches. The current error target of 0.10 eV would make these models more practically useful for researchers' applications. Right: \gls{S2EF} performance as evaluated by mean absolute error of the forces. \gls{IS2RE} and \gls{S2EF} MAEs for their median baselines are 1.756 eV and 0.084 eV/\angstrom, respectively.}
    \label{fig:oc20results}
\end{figure*}

Molecular modeling has progressed at an incredible rate over the past few decades. Simple linear models, neural networks, and kernel methods were originally developed relying on hand-crafted atomic representations,
or descriptors \cite{behler2007generalized,lorenz2004representing,chen2013global,bartok2015g,bartok2010gaussian} as inputs to the models. Descriptors capture invariant geometric
information in the form of bonds and angles of the local environment of an atom. While effective, the parameterization of such descriptors has been a challenging and non-trivial task.  The past few years has seen a shift towards deep learning approaches. Rather than relying on hand crafted representations, models are being developed to learn similar or more expressive representations, specifically by exploiting the graphical nature of molecules using \gls{GNNs} \cite{schutt2017schnet,klicpera2020directional,klicpera2021gemnet,batzner2021se,schutt2021equivariant,liu2021spherical}. Such models only take in 3D atomic coordinates and atomic numbers. A graph is then generated, where atoms are treated as nodes, and the distance between them as edges. Once a graph has been constructed, \gls{GNNs} will undergo several rounds of message passing in which node representations are updated based off messages sent between neighboring nodes. While models may differ in their exact architecture, the update and message functions often include a series of multi-layer perceptrons and nonlinearities. Unlike traditional descriptor based models, \gls{GNNs} end up learning node representations as part of the training process. Learned representations proceed through a final output block where a final prediction is made. In recent years, GNNs have come to surpass traditional descriptor based models \cite{schutt2017schnet,klicpera2020directional,klicpera2021gemnet,batzner2021se,schutt2021equivariant,liu2021spherical}. While typically data hungry, recent models like NequIP \cite{batzner2021se} are demonstrating great performance with as little as 100 samples. \gls{GNNs} continue to gain traction as models continue to demonstrate state of the art performance on molecular datasets.

Since the release of \gls{OC20}, the community has been rapidly developing new approaches to improve existing baselines. Models being developed range from traditional descriptor-style models \cite{lei2021universal} to complex and large GNN architectures \cite{klicpera2021gemnet, shuaibi2021rotation, sriram2021towards, ying2021transformers, godwin2021simple}. Godwin, et al. present a simple, but effective GNN regularization technique to improve graph-level predictions, namely \gls{IS2RE}. Liu, et al. use a similar technique in addition to a graph-based transformer to win $1^{st}$ place in the NeurIPS 2021 Open Catalyst Challenge \cite{ocpchallenge} for direct \gls{IS2RE} predictions. Klicpera, et al.\cite{klicpera2021gemnet, gemnetoc} and Shuaibi, et al.\cite{shuaibi2021rotation} explore various higher order representations (i.e., triplets and quadruplets) and leverage training on the entire \gls{OC20} to achieve impressive performance on the \gls{S2EF} task, with GemNet-OC\cite{gemnetoc} holding the current state of the art across all tasks. Sriram, et al.\cite{sriram2021towards} introduces Graph Parallelism, allowing them to scale GemNet to nearly a billion parameters across multiple GPUs. The scale and diversity of \gls{OC20} has additionally enabled transfer learning approaches to smaller datasets. Kolluru, et al. \cite{kolluru2022transfer} propose a transfer learning technique to use \gls{OC20} pretrained models to improve performance on smaller, out-of-distribution datasets. Similar work has also been demonstrated for other big material datasets \cite{chen2021atomsets}. 

As the community continues to improve performance (Figure \ref{fig:oc20results}), it's important to understand some of the challenges, trends, and pitfalls in developing a generalizable potential.

% OC20 has checkpoints of trained baseline GNNs and since then there are various models added that gives state-of-the-art performance in different tasks. The OC20 paper \cite{chanussot2021open} also talks about the importance of increasing dataset and model size for improving prediction accuracies .
\section{Where are molecular GNNs still erroneous?}

\begin{figure*}
    \centering
    \includegraphics[width=\textwidth]{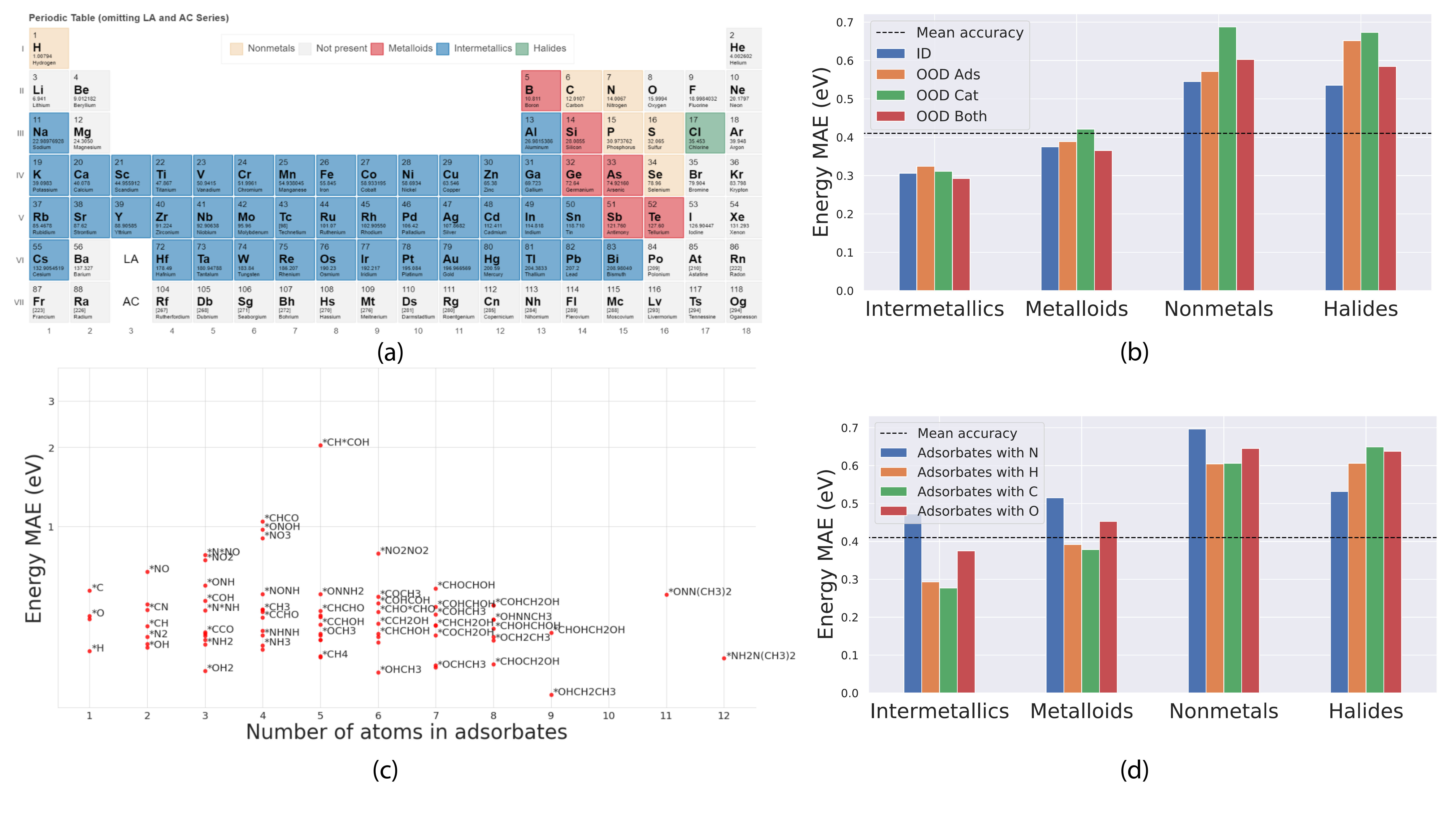}
    \caption{Analysis of GemNet-dT errors on the \gls{OC20} validation sets. (a) The categorization of \gls{OC20} elements into intermetallics, nonmetals, metalloids and halides for analysis. (b) Model performance across the different distributions and material types. (c) Errors averaged across all validation splits for specific adsorbate containing systems. (d) Errors averaged across all validation splits for adsorabtes containing certain elements.}
    \vspace{-4mm}
    \label{fig:errors}
\end{figure*}

Most of the independent work done in developing ML potentials has been confined to datasets built for certain applications. For example, ML potentials for the applications of CO2RR are usually just trained with CO and H adsorbates \cite{back2019convolutional, noh2018active, yoon2020differentiable, chen2020machine}. While this approach might interpolate well across materials, extrapolation to different adsorbates or more complicated materials will likely suffer in performance. A universal ML potential, if possible, would first require a large, diverse dataset that spans material and chemical space. OC20 dataset was created to build ML potentials that cover a large and diverse space of heterogeneous catalysts.

\textbf{Errors across material types:}
With over 300k unique surfaces, \gls{OC20} spans a vast range of material compositions.
When training large \gls{GNNs} on the entire \gls{OC20} dataset, we observe that the accuracies are not uniform across element and adsorbate types. To analyze this, we divide the validation set into four different material types: intermetallics, metalloids, non-metals and halides, Figure \ref{fig:errors}(a). The distribution of data across these classes of materials is not the same, we have significantly more intermetallics and relatively fewer halides. We observe that the performance on non-metals is significantly worse, although both nonmetals and metalloids contribute to similar percentage of training data (Figure \ref{fig:errors}(b)).  On the other hand, models tend to do much better across the board for intermetallics. Inaccuracies coming from non-metals disproportionately contribute to the overall errors, leading to worse performance for both force and energy predictions.

\textbf{Errors across adsorbates:}
Large adsorbates are inherently more complicated as the degrees of freedom increases with the number of atoms. However, we observe no correlation with our model's performance and the size of the adsorbate.  Model accuracies are poor for bidentate adsorbates like *CH*COH, *N*NO, *CH2*O, shown in Figure \ref{fig:errors}(c). Figure \ref{fig:errors}(d) also shows that adsorbates with N and O are generally more erroneous.

% \textbf{Intrinsic DFT errors}

% For all of the ML approaches, we consider DFT calculations using the PBE functional to be the ground truth labels. However, none of the functionals are perfect across materials. 

% However, there could be non-systematic errors or non-systematic cancellation of errors in DFT across systems. This could possibly be an issue since we are trying to fit a single model across the entire dataset. CO puzzle \cite{feibelman2001co, christensen2015identifying}, there are non-systematic errors whenever there are ...

% Bandgaps are not systematically accurate and are functional dependent \cite{perdew1985density, borlido2020exchange, li2013density}.
\begin{table*}[t]
    \centering
    \renewcommand{\arraystretch}{1.0}
    \setlength{\tabcolsep}{5pt}
    \resizebox{0.96\linewidth}{!}{
    \begin{tabular}{lcccccccc}
      \toprule
         \mr{2}{Model} & \mc{4}{c}{Energy MAE (eV) $\downarrow$} &  \mc{4}{c}{Force MAE (eV/\angstrom) $\downarrow$} \\
        & ID & OOD Ads. & OOD Cat. & OOD Both & ID & OOD Ads. & OOD Cat. & OOD Both \\
      \midrule
        Median & 2.04 & 2.42 & 1.99 & 2.58 & 0.081 & 0.080 & 0.079 & 0.098 \\
      \midrule
        & \mc{8}{c}{Gradient forces}\\
        % SpinConv\cite{shuaibi2021rotation}  & 2.8378 & 3.1886 & 2.9021 & 3.7746 & 0.0310 & 0.0350 & 0.0320 & 0.0417 \\
        SpinConv\cite{shuaibi2021rotation}  & - & - & - & - & 0.031 & 0.035 & 0.032 & 0.042 \\
        GemNet-dT\cite{klicpera2021gemnet} & 0.36 & 0.39 & 0.48 & 0.58 & 0.030 & 0.034 & 0.033 & 0.042  \\
     \midrule
        & \mc{8}{c}{Direct forces}\\
        SpinConv\cite{shuaibi2021rotation}  & 0.26 & 0.29 & 0.38 & 0.47 & 0.027 & 0.030 & 0.029 & 0.037 \\
        GemNet-dT\cite{klicpera2021gemnet} & 0.23 & 0.25 & 0.35 & 0.41 & 0.021 & 0.024 & 0.025 & 0.032  \\
    \bottomrule
    \end{tabular}
    }
    \caption{Results on the \gls{OC20} S2EF task via gradient-derived or direct force predictions. All models were trained on the OC20 S2EF All dataset. Results reported for the validation set. Energy metrics are unavailable for the gradient based SpinConv model due to being optimized only on forces.}
    \label{tab:grad}
\end{table*}

\section{Modeling trade-offs}
% \subsection{Gradient or direct force predictions:}
\subsection{Energy-conserving forces}
\setlength{\tabcolsep}{3pt}
\begin{table*}[t]
    \begin{center}
        \resizebox{\textwidth}{!}{
            \begin{tabular}{lcc cccc | cccc  }
            \midrule
            & & & \multicolumn{4}{c|}{Energy MAE [eV] $\downarrow$} & \multicolumn{4}{c}{\gls{EwT} $\uparrow$}  \\
            \cmidrule(l{4pt}r{4pt}){4-7}
            \cmidrule(l{4pt}r{4pt}){8-11}
        Model & Approach & Dataset Size & ID &  OOD Ads & OOD Cat & OOD Both & ID &  OOD Ads & OOD Cat & OOD Both \\
            \midrule
        Median baseline & - & -
                    & $1.75$ & $1.88$ & $1.71$ & $1.66$
                    & $0.71\%$ & $0.72\%$ & $0.89\%$ & $0.74\%$ \\
                \midrule
        DimeNet${+}{+}$~\cite{klicpera2020directional} & Direct & 460,328
                & $0.56$ & $0.73$ & $0.58$ & $0.66$
                    & $4.25\%$ & $2.07\%$ & $4.10\%$ & $2.41\%$ \\
        SpinConv\cite{shuaibi2021rotation} & Direct & 460,328
                    & 0.56   &  0.72   & 0.57 & 0.67
                    & 4.08\%   &  2.26\%   & 3.82\% & 2.33\%\\
                %     GemNet\cite{klicpera2021gemnet} & Direct & 460,328
                % & $$ & $$ & $$ & $$
                %     & $\%$ & $\%$ & $\%$ & $\%$ \\
        NoisyNodes\cite{godwin2021simple} & Direct & 460,328
                    & 0.42  & \textbf{0.57} & 0.44 & \textbf{0.47} & \textbf{9.12\%}  & \textbf{3.49\%}   & 8.01\% & \textbf{4.64\%}\\
        Graphormer\cite{ying2021transformers} & Direct & 460,328
                    &  \textbf{0.40}  &  0.57   & \textbf{0.42} & 0.50 & 8.97\%  &  3.45\%   & \textbf{8.18\%} & 3.79\%\\
                
                \midrule
        DimeNet${+}{+}$ -- LF + LE\cite{klicpera2020directional, chanussot2021open, disc_post} & Relaxation & 2,000,000
                    & $0.53$ & $0.57$ & $0.56$ & $0.52$
                    & $6.79\%$ & $4.71\%$ & $6.49\%$ & $4.54\%$ \\
        SpinConv\cite{shuaibi2021rotation, disc_post}  & Relaxation & 2,000,000
                & $0.46$ & $0.51$ & $0.47$ & $0.44$
                    & $7.38\%$ & $4.82\%$ & $7.05\%$ & $5.31\%$\\
        GemNet-dT\cite{klicpera2021gemnet} & Relaxation & 2,000,000
                & $0.44$ & $0.44$ & $0.45$ & $0.42$
                    & $9.37\%$ & $6.59\%$ & $8.42\%$ & $6.40\%$\\
        GemNet-OC\cite{gemnetoc} & Relaxation & 2,000,000&
            $\textbf{0.41}$  & $\textbf{0.42}$  & $\textbf{0.42}$  & $\textbf{0.39}$  &
            $\textbf{11.02\%}$ & $\textbf{8.68\%}$  & $\textbf{10.10\%}$ & $\textbf{7.82\%}$\\
        \midrule
        DimeNet${+}{+}$ -- LF + LE \cite{klicpera2020directional, chanussot2021open} & Relaxation & 133,934,018
                    & $0.50$ & $0.54$ & $0.58$ & $0.61$
                    & $6.57\%$ & $4.34\%$ & $5.09\%$ & $3.93\%$ \\
        SpinConv\cite{shuaibi2021rotation}  & Relaxation & 133,934,018
                & $0.42$ & $0.44$ & $0.46$ & $0.42$
                    & $9.37\%$ & $7.47\%$ & $8.16\%$ & $6.56\%$\\
        GemNet-dT\cite{klicpera2021gemnet} & Relaxation & 133,934,018
                & 0.39 & 0.39 & 0.43 & 0.38 & $12.37\%$ & $9.11\%$ & $10.09\%$ & $7.87\%$ \\
        GemNet-OC\cite{gemnetoc} & Relaxation & 133,934,018
                & \textbf{0.35} & \textbf{0.35} & \textbf{0.38} & \textbf{0.34} & $\textbf{16.06\%}$ & $\textbf{12.62\%}$ & $\textbf{13.17\%}$ & $\textbf{11.06\%}$ \\
            \bottomrule
            \end{tabular}}
    \end{center}
    \caption{Results on the OC20 \gls{IS2RE} task using one of two approaches. \textbf{Direct} Directly predicting the relaxed state energy and \textbf{Relaxation} Training a model for energy and force predictions, followed by an iterative ML-based geometry optimization to arrive at a relaxed structure and energy. Relaxation results on the 2M subset suggest that competitive results are still possible with a limited compute budget. Results reported for the test set.}
    \label{tab:is2re}
\end{table*}
Force predictions play an important role in the applications of ML models for catalyst discovery. While some tasks may only be interested in property predictions like adsorption or formation energy \cite{back2019convolutional, li2017high, mamun2019high}, forces are necessary to study dynamics such as structural relaxations, molecular dynamics, and transition state calculations \cite{chanussot2021open, batzner2021se, schutt2017schnet, peterson2016acceleration}.

Physically, energy-conserving forces are derived as the gradient of energy with respect to atomic positions:
\begin{equation}
    \boldsymbol{F_i} = -\frac{dE}{dx_i}
\end{equation}

Energy-conservation is critical in studying molecular dynamics accurately. ML models estimating energy-conserving forces must ensure the architecture is continuous and differentiable, often satisfied by appropriate non-linear activation functions \cite{schutt2017schnet, klicpera2020directional, klicpera2021gemnet}. Geometrically, forces derived in an energy-conserving manner ensures forces are rotationally equivariant, a necessary physical relation of molecular systems \cite{christensen2020role}. Unfortunately, a gradient calculation increases model overhead in both memory usage and computational time by a factor of 2-4 \cite{shuaibi2021rotation, hu2021forcenet}. For datasets like MD17, calculating forces as a gradient is known to help in model accuracies as that is an important physical prior to the model \cite{batzner2021se, klicpera2021gemnet, shuaibi2021rotation}. Models trained on MD17 are often used to run molecular dynamics, further necessitating the need for energy-conservation \cite{batzner2021se}. However, for the \gls{OC20} dataset, particularly in the task of geometric optimization, we observe that the gradient approach for calculating forces to perform worse than direct prediction of forces for GemNet-dT \cite{klicpera2021gemnet} and Spinconv \cite{shuaibi2021rotation}. Dimenet ++ \cite{klicpera2020directional} and ForceNet \cite{hu2021forcenet} were built for gradient and direct approach respectively. The gradient approach could also make the training unstable in certain cases, which  has been observed for ForceNet\cite{hu2021forcenet} and GemNet-OC\cite{gemnetoc}. Table \ref{tab:grad} compares performance on the S2EF task for two recent top performing models, GemNet-dT \cite{klicpera2021gemnet} and SpinConv \cite{shuaibi2021rotation}. Not only are the force accuracies worse for the gradient approach, but the corresponding relaxed structure and relaxed energy metrics calculated via optimization are also significantly worse \cite{shuaibi2021rotation}.

While energy-conservation plays a critical role in many molecular applications, we observe that direct force computations brings efficiency and performance advantages \cite{hu2021forcenet, shuaibi2021rotation}. Models trained for direct force predictions are limited to applications where strict enforcement of energy-conservation can reasonably be ignored, i.e. \gls{OC20}'s structural relaxations. Here, atomic positions are updated solely from force estimates \cite{chanussot2021open, del2019local}. If necessary, DFT, or a subsequent ML model, can then be used to make reliable energy predictions on the ML optimized structure. Similarly, transition states or saddle points can be derived in a similar manner with direct-force models. We want to emphasize that although unorthodox, direct-force models still prove to be useful in certain catalyst applications, i.e. \gls{OC20}-like tasks.

\subsection{Prediction of relaxed energy and structure}

Adsorption energy is one of many properties that helps inform catalyst performance\cite{wang2018adsorption}. Computationally, this is computed via a series of QM structural relaxations. The relaxed energy is then referenced to represent the adsorption energy, see Chanussot et al.\cite{chanussot2021open}, Garc{\'\i}a-Muelas et al.\cite{garcia2019statistical} for more details. From a data-driven approach, we can predict the relaxed energy or the relaxed structure of an atomic system usually via two methods. First, we can build a surrogate to DFT, approximating system energy and per-atom structures, and running ML optimizations to find the minimum energy, a common approach within the field. Alternatively, given a large enough dataset of relaxed structures and energies, we can try to predict these properties directly using a ML model instead of optimizing via an iterative loop. The advantage of the direct method over the relaxation approach is that it requires only a single call to the ML model, whereas the relaxation approach could require on average 200-300 calls for a single relaxation. Direct approaches are particularly advantageous when we talk about the computational cost of approaching large scale inference on the order of hundreds of millions to billions of systems.

The community has made tremendous progress in predicting adsorption energy as evaluated by the \gls{OC20} IS2RE task (Figure \ref{fig:oc20results}). Direct approaches, despite using ~300x less data, are approaching the competitive relaxation based approaches of GemNet-XL and GemNet-OC. Inference time aside, models trained on the full 133M dataset for the relaxation based approaches are typically compute intensive, using between 128-512 GPUs \cite{shuaibi2021rotation, hu2021forcenet, klicpera2021gemnet, chanussot2021open}. While this is certainly a small price to pay if the models developed accelerate the discovery process, it does make it difficult for the community to engage in and aid in development. This has been particularly observed in the NeurIPS 2021 Open Catalyst Challenge \cite{ocpchallenge}, where of the 30 submissions, 0 were made via the relaxation approach. Here, we show that models trained on a 2M subset of the full dataset are still able to provide competitive results and even, averaged across all splits, out perform direct approaches. Given the trends in the 2M dataset correlate well with the full 133M dataset \cite{gemnetoc}, this should help incentivize the community to explore other approaches even with resource limitations. Although the relaxation approach is computationally expensive for both training and inference, we have observed that the models trained through this approach tend to generalize better on out-of-distribution (OOD) data, Table \ref{tab:is2re}.

Direct relaxed energy predictions are an easier ML problem than direct structure predictions. For a system of size N, energy predictions require a single scalar output, while structure predictions require 3N components. We find that for relaxed energy prediction tasks, metrics are closer for direct and relaxation approach whereas for structure prediction task the metrics are worse. The OC20 paper provides a baseline for relaxed structure prediction only via the relaxation approach \cite{chanussot2021open}. In Table \ref{tab:is2rs} we provide baselines for direct relaxed structure prediction. A considerable gap exists between the direct and relaxation based approaches (especially in the DFT based metrics).

\subsection{Metrics for finding local minima}

Relaxed structure prediction is less straightforward than some of the other common energy and force prediction tasks. Given a dataset like \gls{OC20} where relaxed structures are not necessarily global minima, a model trained on such a dataset could either (1) predict and arrive at the same local minima, (2) arrive at a different, but still suitable minima, or (3) fail to arrive at any sort of minima.

To account for this, two main metrics have been presented in the \gls{OC20} paper. \gls{ADwT} is a distance based metric and measures how close the predicted structure compares to the actual structure. This is similar to the Global Distance Test (GDT) metric in the protein folding task \cite{jumper2021highly, zemla2003lga}. \gls{ADwT} takes an average across different thresholds varying from 0.1 to 0.5\angstrom to ensure a signal is captured. For the \gls{OC20} dataset, we evaluate this metric for the input initial structures for an accuracy of 21.18\% on the in-domain validation set \cite{chanussot2021open}. Models, at the bare minimum, should perform better than this baseline. To ensure invariance to arbitrary coordinate reference frames, we predict the difference between initial and final positions instead of the final position Cartesian coordinates. Predicting the delta difference helps simplify this task and results in improved \gls{ADwT} accuracies.

\begin{table}
    \centering
    \caption{Baseline metrics for IS2RS direct task in comparison with the relaxation approach. Metrics are reported on a 2k subset of the validation set, across all splits. DwT is evaluated at a threshold of 0.04 \AA. For compute reasons, DFT-based metrics were evaluated on a 200 system subset of the 2k, 50 systems from each split.} 
    \label{tab:is2rs}
    \resizebox{\columnwidth}{!}{
    \begin{tabular}{llcccc}
      \hline
        & Model & DwT (at 0.04 \angstrom) $\uparrow$ & ADwT $\uparrow$ & FbT* $\uparrow$ & AFbT* $\uparrow$ \\
        \hline
\multirow{3}{*}{Direct} &             ForceNet\cite{hu2021forcenet} &
        0.70 & 45.69\%& 0.00\%& 0.00\%\\
        & SpinConv\cite{shuaibi2021rotation} &
        1.05 & 47.76\%& 0.00\% & 0.00\%\\
        & GemNet-dT\cite{klicpera2021gemnet} &
        1.75 & 45.87\%& 0.00\% & 0.08\%\\
        \midrule
\multirow{3}{*}{Relaxation} &             ForceNet\cite{hu2021forcenet} &
        1.45 & 46.51\% & 0.00\% & 7.64\%\\
        & SpinConv\cite{shuaibi2021rotation} &
        8.20 & 55.81\%& 0.00\% & 12.55\%\\
        & GemNet-dT\cite{klicpera2021gemnet} &
        13.95 & 60.88\% & 0.00\% & 20.35\%\\
      \bottomrule
    \end{tabular}}
\end{table}

A model that predicts a relaxed structure that is not identical to its DFT reference may still be considered successful for two reasons. (1) the model could have predicted a symmetrically identical site on the surface and (2) the model predicted a different, but still suitable local minima. The former is more a concern surrounding the distance-based metric, as \gls{ADwT}, although accounts for periodic-boundary conditions, does not consider symmetrically identical sites. While it is rather unlikely an adsorbate initialized over a particular site will hop several sites over to a symmetrically identical site, it is worth raising awareness to the possibility. On the other hand, a model that arrives at a different relaxed structure entirely will fail according to \gls{ADwT}. However, to verify whether the model has predicted a different suitable minima, we can evaluate the DFT forces corresponding to the ML predicted structures. This metric is called \gls{AFbT} and it measures the percent of structures having their forces close to zero \cite{chanussot2021open}. Since models are expected to predict relaxed structures, DFT forces should be close to zero. This is a stricter metric as compared to \gls{ADwT}. However, this is far more expensive due to the additional DFT calculations. A more practically useful metric would be number of DFT calculations required to find the relaxed structure starting from the ML relaxed structure. This would give us an idea of the percent of DFT calculations that the current ML models can reduce. Although useful, this is a significantly more expensive metric than \gls{AFbT} calculations. While it is not something \gls{OCP} tracks on their public leaderboard, we bring awareness to it as there could be instances where models do poorly on \gls{ADwT} and \gls{AFbT} but resulting structures are only a few \gls{DFT} steps away from the relaxed structure.

In Table \ref{tab:is2rs} we compare relaxed structure prediction via a direct and relaxation approach. We observe that direct methods, although having competitive \gls{ADwT} metrics, have \gls{AFbT} metrics that are significantly worse. This suggests that direct models do a reasonable job at getting close to the relaxed structure but are in high-force configurations, failing to capture repulsive physical interactions \cite{godwin2021simple}. We speculate models struggle with this since small perturbations distances can have large consequences on forces, e.g. moving two atoms at an equilibrium bond length fractions of an angstrom towards each other. Relaxed structure prediction via the relaxation approach avoids this issue by using ML forces to drive a geometric optimizer.

We observe that distance metrics at tighter thresholds correlate better with force based metrics, however, going below 0.04 $\angstrom$ does not give sufficient signal and the accuracies for most systems fall to zero. Moreover, the \gls{DwT} at 0.04 $\angstrom$ isn't a good enough signal that can replace \gls{AFbT}. For example, DwT (at 0.04\angstrom) for ForceNet relaxation approach and GemNet-dT direct approach are similar, however, the AFbT metrics still differ by 7.56\% (as shown in Table \ref{tab:is2rs}).  We believe that finding non DFT-based metrics that correlate well with DFT-based metrics is still an open and important question in the community which would make model evaluation computationally less expensive.

\subsection{Additional data}

\noindent The OC20 paper~\cite{chanussot2021open} released two additional data subsets
generated with ab-initio molecular dynamics (`MD') and
structural perturbations (`Rattled').
These provide $38M$ and $17M$ additional \gls{S2EF} training data points
respectively.

Table~\ref{tab:gnoc_rattled_md} presents results for
GemNet-OC\cite{gemnetoc} models trained on \gls{S2EF}, Rattled, and MD data compared
against similar analysis from the OC20 paper for
DimeNet++~\cite{klicpera2020directional,klicpera2020fast}.
First, on the force MAE metric, addition of MD data hurts DimeNet++ while it
improves GemNet-OC. We speculate this to be another artifact of modeling
forces as negative gradients of energy (as in DimeNet++) \textit{vs}. direct
prediction (as in GemNet-OC).
Second, consistent with the OC20 paper, adding MD data to the training set provides
a useful signal for \gls{IS2RS} structure relaxations as per the AFbT metric.
Finally, adding Rattled data helps with \gls{IS2RS} metrics, but did not help or marginally hurt the \gls{S2EF} force MAE.
This could be due to a variety of reasons -- random perturbations being too
large / small to be useful, intermediate structures along a
trajectory being less useful compared to closer to the local minimum (as in MD initial structures), \text{etc}.
A promising direction here could be active learning approaches to optimally query
additional training data points.

\setlength{\tabcolsep}{3pt}
\begin{table}[h!]
    \begin{center}
        \resizebox{0.47\textwidth}{!}{
            \begin{tabular}{cl ccc }
                \toprule
                & & \multicolumn{1}{c}{S2EF Val ID} & \multicolumn{2}{c}{IS2RS Test} \\
                \cmidrule(l{4pt}r{4pt}){3-3}
                \cmidrule(l{4pt}r{4pt}){4-5}
                & Training Data (\# samples) &
                    Force MAE $\downarrow$ & ADwT $\uparrow$ & AFbT $\uparrow$ \\
                \midrule
                \multirow{3}{*}{\rotatebox[origin=c]{90}{DN++} $\begin{dcases} \\ \\ \end{dcases}$}
                & $20M$ ($20M$) &
                     $0.0511$ & $34.37\%$ & $2.67\%$ \\
                & $20M$ $+$ MD ($58M$) &
                     $0.0594$ & $47.69\%$ & $17.09\%$ \\
                & $20M$ $+$ Rattled ($37M$) &
                     $0.0614$ & $43.94\%$ & $12.51\%$ \\
                \midrule
                \multirow{3}{*}{\rotatebox[origin=c]{90}{GN-OC} $\begin{dcases} \\ \\ \end{dcases}$}
                & All ($133M$) &
                    $0.0179$ & $60.33\%$ & $35.27\%$ \\
                & All $+$ MD ($172M$) &
                    $0.0173$ & $60.77\%$ & $38.05\%$ \\
                & All $+$ MD $+$ Rattled ($189M$) &
                    $0.0174$ & - & - \\
                \bottomrule
            \end{tabular}}
    \end{center}
    \caption{Results with DimeNet++ (DN++) and GemNet-OC (GN-OC) trained on MD and Rattled. S2EF results reported for the validation in-distribution set. IS2RS results reported on the test set. }
    \label{tab:gnoc_rattled_md}
\end{table}
\setlength{\tabcolsep}{1.4pt}

\section{Summary and Outlook}

The development of generalizable or universal ML models has only recently been seriously considered with the emergence of large-scale datasets like \gls{OC20} \cite{chanussot2021open}. Since its release, the catalysis and ML communities have both made tremendous progressive in developing models for catalyst applications. As the community continues to grow and as more datasets emerge that span material and composition space, the prospect of large-scale generalizable models is within reason. Progress thus far has demonstrated several challenges in accomplishing this feat: classes of materials and adsorbates with inconsistent errors, energy-conserving forces, relaxed vs direct approaches, DFT metrics, and data augmentation strategies. In this perspective, we discussed these challenges in detail and provided some insights as to how and why they are important. Although these challenges were discussed in the context of \gls{OC20}, we anticipate similar challenges to future datasets of its kind.

Datasets like \gls{OC20} has offered new ways to how we think about building large, generalizable, and reliable models. While model performance has been the focal point of community progress thus far, we provide an outlook of other important challenges that we hope the community to engage in.

\textbf{Training strategies}. \gls{OC20} was released with predefined training, validation, and test sets. Its splits were curated in a manner to tackle the problem of building a single generalizable model for catalysis. However, it could be the case that multiple models for different subsets of the data, e.g. adsorbates, compositions, materials, do better. In the case of nonmetals, for instance, we have shown that this actually hurts performance - a possible consequence of the reduced dataset size.

\textbf{Uncertainty and active learning}. While model performance is a necessary step for the discovery process, it is not always sufficient. A practical ML-aided catalyst discovery pipeline will ultimately turn to experiments to validate whether the ML predicted ``great" catalyst is at all effective. Having confidence in these predictions is particularly important to avoid wasted expensive experiments. Uncertainty quantification has been a particularly popular topic within the catalysis community, often focused on the small data regime and active learning \cite{tran2020methods, busk2021calibrated, shuaibi2020enabling, vandermause2020fly, jinnouchi2019phase, torres2019low, scalia2020evaluating}. The effectiveness of traditional uncertainty estimation techniques on large datasets like \gls{OC20} is a necessary and important step for the future of this work. Similarly, how to best leverage active learning for either dataset generation and/or augmentation \cite{smith2020ani} or online active learning \cite{vandermause2020fly, shuaibi2020enabling} at the scale of \gls{OC20} will be an exciting future direction.

\textbf{Model efficiency}. In addition to model performance and reliability, model efficiency will continue to be critical for all applications. For training, faster, more data efficient models can help attract the community to tackle some of the bigger challenges like a surrogate to DFT, i.e. \gls{OC20}'s S2EF task. Progress so far has shown that the best models are also the largest models. From an inference perspective, this poses obvious challenges of slower speeds and ultimately reduced screening throughput. While models still remain orders of magnitude faster than DFT, when considering the possibility of screening billions of systems, computational costs add up. Recent models encoding equivariant representations \cite{batzner2021se, musaelian2022learning} have shown incredible scaling and efficiency gains that could be promising to explore. Moving forward, efficient architectures and model distillation\cite{frankle2018lottery} will be an important contribution to reduce the computational cost of large-scale inference, even if it means sacrificing some accuracy.

\textbf{Data augmentation}. The scale of \gls{OC20} makes data augmentation a non-trivial challenge. With 130M+ training data points, randomly adding 10-100k data points will likely have negligible impact on the models. We observed that models using the additional MD data are able to perform the best, while the rattled data has little impact. Identifying strategies to combine and train large molecular and material datasets like ANI-1\cite{smith2017ani} and OQMD\cite{kirklin2015open} with \gls{OC20} could help improve models even further. The biggest challenge surrounding this comes from combining datasets of varying levels of DFT theory.

% remove?
\textbf{Energy-conserving forces}. In the context of \gls{OC20}, we have observed that the best performing models make a direct force-prediction. While this may be suitable for some applications, the more physically motivated gradient approach to force prediction is desired for other applications like MD. The same direct models applied to MD17 observe the opposite effect, better performance via the gradient method \cite{gemnetoc}. It remains an open question why this is the case, and we encourage others to investigate this observation.

\textbf{Physics-based modeling}. The majority of models submitted to \gls{OC20} have followed a purely data-driven approach, only taking in atomic numbers and positions as inputs. Exploring ways to leverage \gls{OC20} charge density or Bader charge data \footnote{To be made publically available at https://github.com/Open-Catalyst-Project/ocp/blob/main/DATASET.md} could prove useful, particularly in the low data regime. Additionally, models like UNiTE \cite{qiao2021unite} or OrbNet\cite{qiao2020orbnet} that leverage tight binding DFT\cite{bannwarth2019gfn2} for featurization could be interesting to explore for catalyst applications.

% \begin{acknowledgement}

% We thank Johannes Gasteiger for helpful discussions on the challenges of energy-conserving forces.
% Please use ``The authors thank \ldots'' rather than ``The
% authors would like to thank \ldots''.

% The author thanks Mats Dahlgren for version one of \textsf{achemso},
% and Donald Arseneau for the code taken from \textsf{cite} to move
% citations after punctuation. Many users have provided feedback on the
% class, which is reflected in all of the different demonstrations
% shown in this document.

% \end{acknowledgement}
\clearpage 

\printglossary[title=Glossary, type=\acronymtype]
\printglossary

\clearpage

% \begin{suppinfo}

% This will usually read something like: ``Experimental procedures and
% characterization data for all new compounds. The class will
% automatically add a sentence pointing to the information on-line:

% \end{suppinfo}

\bibliography{main}

\providecommand{\latin}[1]{#1}
\makeatletter
\providecommand{\doi}
  {\begingroup\let\do\@makeother\dospecials
  \catcode`\{=1 \catcode`\}=2 \doi@aux}
\providecommand{\doi@aux}[1]{\endgroup\texttt{#1}}
\makeatother
\providecommand*\mcitethebibliography{\thebibliography}
\csname @ifundefined\endcsname{endmcitethebibliography}
  {\let\endmcitethebibliography\endthebibliography}{}
\begin{mcitethebibliography}{76}
\providecommand*\natexlab[1]{#1}
\providecommand*\mciteSetBstSublistMode[1]{}
\providecommand*\mciteSetBstMaxWidthForm[2]{}
\providecommand*\mciteBstWouldAddEndPuncttrue
  {\def\EndOfBibitem{\unskip.}}
\providecommand*\mciteBstWouldAddEndPunctfalse
  {\let\EndOfBibitem\relax}
\providecommand*\mciteSetBstMidEndSepPunct[3]{}
\providecommand*\mciteSetBstSublistLabelBeginEnd[3]{}
\providecommand*\EndOfBibitem{}
\mciteSetBstSublistMode{f}
\mciteSetBstMaxWidthForm{subitem}{(\alph{mcitesubitemcount})}
\mciteSetBstSublistLabelBeginEnd
  {\mcitemaxwidthsubitemform\space}
  {\relax}
  {\relax}

\bibitem[Chanussot \latin{et~al.}(2021)Chanussot, Das, Goyal, Lavril, Shuaibi,
  Riviere, Tran, Heras-Domingo, Ho, Hu, \latin{et~al.}
  others]{chanussot2021open}
Chanussot,~L.; Das,~A.; Goyal,~S.; Lavril,~T.; Shuaibi,~M.; Riviere,~M.;
  Tran,~K.; Heras-Domingo,~J.; Ho,~C.; Hu,~W., \latin{et~al.}  Open Catalyst
  2020 (OC20) Dataset and Community Challenges. \emph{ACS Catalysis}
  \textbf{2021}, \emph{11}, 6059--6072\relax
\mciteBstWouldAddEndPuncttrue
\mciteSetBstMidEndSepPunct{\mcitedefaultmidpunct}
{\mcitedefaultendpunct}{\mcitedefaultseppunct}\relax
\EndOfBibitem
\bibitem[Ren \latin{et~al.}(2018)Ren, Ward, Williams, Laws, Wolverton,
  Hattrick-Simpers, and Mehta]{ren2018accelerated}
Ren,~F.; Ward,~L.; Williams,~T.; Laws,~K.~J.; Wolverton,~C.;
  Hattrick-Simpers,~J.; Mehta,~A. Accelerated discovery of metallic glasses
  through iteration of machine learning and high-throughput experiments.
  \emph{Science advances} \textbf{2018}, \emph{4}, eaaq1566\relax
\mciteBstWouldAddEndPuncttrue
\mciteSetBstMidEndSepPunct{\mcitedefaultmidpunct}
{\mcitedefaultendpunct}{\mcitedefaultseppunct}\relax
\EndOfBibitem
\bibitem[Jose \latin{et~al.}(2006)Jose, Zhanpeisov, Fukumura, Baba, and
  Ishikawa]{jose2006structure}
Jose,~R.; Zhanpeisov,~N.~U.; Fukumura,~H.; Baba,~Y.; Ishikawa,~M. Structure-
  property correlation of CdSe clusters using experimental results and
  first-principles DFT calculations. \emph{Journal of the American Chemical
  Society} \textbf{2006}, \emph{128}, 629--636\relax
\mciteBstWouldAddEndPuncttrue
\mciteSetBstMidEndSepPunct{\mcitedefaultmidpunct}
{\mcitedefaultendpunct}{\mcitedefaultseppunct}\relax
\EndOfBibitem
\bibitem[Rahali \latin{et~al.}(2021)Rahali, Ben~Aissa, Khezami, Elamin, Seydou,
  and Modwi]{rahali2021adsorption}
Rahali,~S.; Ben~Aissa,~M.~A.; Khezami,~L.; Elamin,~N.; Seydou,~M.; Modwi,~A.
  Adsorption behavior of Congo red onto barium-doped ZnO nanoparticles:
  correlation between experimental results and DFT calculations.
  \emph{Langmuir} \textbf{2021}, \emph{37}, 7285--7294\relax
\mciteBstWouldAddEndPuncttrue
\mciteSetBstMidEndSepPunct{\mcitedefaultmidpunct}
{\mcitedefaultendpunct}{\mcitedefaultseppunct}\relax
\EndOfBibitem
\bibitem[Xu \latin{et~al.}(2019)Xu, Meng, Li, Li, Sui, Wang, and
  Yang]{xu2019dissolution}
Xu,~L.; Meng,~X.; Li,~M.; Li,~W.; Sui,~Z.; Wang,~J.; Yang,~J. Dissolution and
  degradation of nuclear grade cationic exchange resin by Fenton oxidation
  combining experimental results and DFT calculations. \emph{Chemical
  Engineering Journal} \textbf{2019}, \emph{361}, 1511--1523\relax
\mciteBstWouldAddEndPuncttrue
\mciteSetBstMidEndSepPunct{\mcitedefaultmidpunct}
{\mcitedefaultendpunct}{\mcitedefaultseppunct}\relax
\EndOfBibitem
\bibitem[Ktari \latin{et~al.}(2015)Ktari, Fourati, Zerrouki, Ruan, Seydou,
  Barbaut, Nal, Yaakoubi, Chehimi, and Kalfat]{ktari2015design}
Ktari,~N.; Fourati,~N.; Zerrouki,~C.; Ruan,~M.; Seydou,~M.; Barbaut,~F.;
  Nal,~F.; Yaakoubi,~N.; Chehimi,~M.; Kalfat,~R. Design of a polypyrrole
  MIP-SAW sensor for selective detection of flumequine in aqueous media.
  Correlation between experimental results and DFT calculations. \emph{RSC
  advances} \textbf{2015}, \emph{5}, 88666--88674\relax
\mciteBstWouldAddEndPuncttrue
\mciteSetBstMidEndSepPunct{\mcitedefaultmidpunct}
{\mcitedefaultendpunct}{\mcitedefaultseppunct}\relax
\EndOfBibitem
\bibitem[Rode \latin{et~al.}(2018)Rode, Abdalghani, Arcadi, Aschi, Chiarini,
  and Marinelli]{rode2018synthesis}
Rode,~N.~D.; Abdalghani,~I.; Arcadi,~A.; Aschi,~M.; Chiarini,~M.; Marinelli,~F.
  Synthesis of 2-Acylindoles via Ag-and Cu-Catalyzed anti-Michael
  Hydroamination of $\beta$-(2-Aminophenyl)-$\alpha$, $\beta$-ynones:
  Experimental Results and DFT Calculations. \emph{The Journal of organic
  chemistry} \textbf{2018}, \emph{83}, 6354--6362\relax
\mciteBstWouldAddEndPuncttrue
\mciteSetBstMidEndSepPunct{\mcitedefaultmidpunct}
{\mcitedefaultendpunct}{\mcitedefaultseppunct}\relax
\EndOfBibitem
\bibitem[Pause \latin{et~al.}(2001)Pause, Robert, Heinicke, and
  K{\"u}hl]{pause2001radical}
Pause,~L.; Robert,~M.; Heinicke,~J.; K{\"u}hl,~O. Radical anions of carbenes
  and carbene homologues. DFT study and preliminary experimental results.
  \emph{Journal of the Chemical Society, Perkin Transactions 2} \textbf{2001},
  1383--1388\relax
\mciteBstWouldAddEndPuncttrue
\mciteSetBstMidEndSepPunct{\mcitedefaultmidpunct}
{\mcitedefaultendpunct}{\mcitedefaultseppunct}\relax
\EndOfBibitem
\bibitem[Agrawal and Choudhary(2016)Agrawal, and
  Choudhary]{agrawal2016perspective}
Agrawal,~A.; Choudhary,~A. Perspective: Materials informatics and big data:
  Realization of the “fourth paradigm” of science in materials science.
  \emph{Apl Materials} \textbf{2016}, \emph{4}, 053208\relax
\mciteBstWouldAddEndPuncttrue
\mciteSetBstMidEndSepPunct{\mcitedefaultmidpunct}
{\mcitedefaultendpunct}{\mcitedefaultseppunct}\relax
\EndOfBibitem
\bibitem[Guan \latin{et~al.}(2022)Guan, Chaffart, Liu, Tan, Zhang, Wang, Li,
  and Ricardez-Sandoval]{guan2022machine}
Guan,~Y.; Chaffart,~D.; Liu,~G.; Tan,~Z.; Zhang,~D.; Wang,~Y.; Li,~J.;
  Ricardez-Sandoval,~L. Machine learning in solid heterogeneous catalysis:
  Recent developments, challenges and perspectives. \emph{Chemical Engineering
  Science} \textbf{2022}, \emph{248}, 117224\relax
\mciteBstWouldAddEndPuncttrue
\mciteSetBstMidEndSepPunct{\mcitedefaultmidpunct}
{\mcitedefaultendpunct}{\mcitedefaultseppunct}\relax
\EndOfBibitem
\bibitem[Rosen \latin{et~al.}(2022)Rosen, Notestein, and
  Snurr]{rosen2022realizing}
Rosen,~A.~S.; Notestein,~J.~M.; Snurr,~R.~Q. Realizing the data-driven,
  computational discovery of metal-organic framework catalysts. \emph{Current
  Opinion in Chemical Engineering} \textbf{2022}, \emph{35}, 100760\relax
\mciteBstWouldAddEndPuncttrue
\mciteSetBstMidEndSepPunct{\mcitedefaultmidpunct}
{\mcitedefaultendpunct}{\mcitedefaultseppunct}\relax
\EndOfBibitem
\bibitem[Himanen \latin{et~al.}(2019)Himanen, Geurts, Foster, and
  Rinke]{himanen2019data}
Himanen,~L.; Geurts,~A.; Foster,~A.~S.; Rinke,~P. Data-driven materials
  science: status, challenges, and perspectives. \emph{Advanced Science}
  \textbf{2019}, \emph{6}, 1900808\relax
\mciteBstWouldAddEndPuncttrue
\mciteSetBstMidEndSepPunct{\mcitedefaultmidpunct}
{\mcitedefaultendpunct}{\mcitedefaultseppunct}\relax
\EndOfBibitem
\bibitem[Xu \latin{et~al.}(2021)Xu, Cao, and Hu]{xu2021perspective}
Xu,~J.; Cao,~X.-M.; Hu,~P. Perspective on computational reaction prediction
  using machine learning methods in heterogeneous catalysis. \emph{Physical
  Chemistry Chemical Physics} \textbf{2021}, \emph{23}, 11155--11179\relax
\mciteBstWouldAddEndPuncttrue
\mciteSetBstMidEndSepPunct{\mcitedefaultmidpunct}
{\mcitedefaultendpunct}{\mcitedefaultseppunct}\relax
\EndOfBibitem
\bibitem[Tran and Ulissi(2018)Tran, and Ulissi]{tran2018active}
Tran,~K.; Ulissi,~Z.~W. Active learning across intermetallics to guide
  discovery of electrocatalysts for \ce{CO2} reduction and \ce{H2} evolution.
  \emph{Nature Catalysis} \textbf{2018}, \emph{1}, 696--703\relax
\mciteBstWouldAddEndPuncttrue
\mciteSetBstMidEndSepPunct{\mcitedefaultmidpunct}
{\mcitedefaultendpunct}{\mcitedefaultseppunct}\relax
\EndOfBibitem
\bibitem[Back \latin{et~al.}(2019)Back, Tran, and Ulissi]{back2019toward}
Back,~S.; Tran,~K.; Ulissi,~Z.~W. Toward a design of active oxygen evolution
  catalysts: insights from automated density functional theory calculations and
  machine learning. \emph{Acs Catalysis} \textbf{2019}, \emph{9},
  7651--7659\relax
\mciteBstWouldAddEndPuncttrue
\mciteSetBstMidEndSepPunct{\mcitedefaultmidpunct}
{\mcitedefaultendpunct}{\mcitedefaultseppunct}\relax
\EndOfBibitem
\bibitem[Ying \latin{et~al.}(2021)Ying, Fan, Luo, Qiao, and
  Huang]{ying2021unravelling}
Ying,~Y.; Fan,~K.; Luo,~X.; Qiao,~J.; Huang,~H. Unravelling the origin of
  bifunctional OER/ORR activity for single-atom catalysts supported on C2N by
  DFT and machine learning. \emph{Journal of Materials Chemistry A}
  \textbf{2021}, \emph{9}, 16860--16867\relax
\mciteBstWouldAddEndPuncttrue
\mciteSetBstMidEndSepPunct{\mcitedefaultmidpunct}
{\mcitedefaultendpunct}{\mcitedefaultseppunct}\relax
\EndOfBibitem
\bibitem[Ge \latin{et~al.}(2020)Ge, Yuan, Min, Li, Chen, Xu, and
  Goddard~III]{ge2020predicted}
Ge,~L.; Yuan,~H.; Min,~Y.; Li,~L.; Chen,~S.; Xu,~L.; Goddard~III,~W.~A.
  Predicted optimal bifunctional electrocatalysts for the hydrogen evolution
  reaction and the oxygen evolution reaction using chalcogenide
  heterostructures based on machine learning analysis of in silico quantum
  mechanics based high throughput screening. \emph{The Journal of Physical
  Chemistry Letters} \textbf{2020}, \emph{11}, 869--876\relax
\mciteBstWouldAddEndPuncttrue
\mciteSetBstMidEndSepPunct{\mcitedefaultmidpunct}
{\mcitedefaultendpunct}{\mcitedefaultseppunct}\relax
\EndOfBibitem
\bibitem[Toniato \latin{et~al.}(2022)Toniato, Vaucher, and
  Laino]{toniato2022grand}
Toniato,~A.; Vaucher,~A.~C.; Laino,~T. Grand challenges on accelerating
  discovery in catalysis. \emph{Catalysis Today} \textbf{2022}, \emph{387},
  140--142\relax
\mciteBstWouldAddEndPuncttrue
\mciteSetBstMidEndSepPunct{\mcitedefaultmidpunct}
{\mcitedefaultendpunct}{\mcitedefaultseppunct}\relax
\EndOfBibitem
\bibitem[Chmiela \latin{et~al.}(2017)Chmiela, Tkatchenko, Sauceda, Poltavsky,
  Sch{\"u}tt, and M{\"u}ller]{chmiela2017machine}
Chmiela,~S.; Tkatchenko,~A.; Sauceda,~H.~E.; Poltavsky,~I.; Sch{\"u}tt,~K.~T.;
  M{\"u}ller,~K.-R. Machine learning of accurate energy-conserving molecular
  force fields. \emph{Science advances} \textbf{2017}, \emph{3}, e1603015\relax
\mciteBstWouldAddEndPuncttrue
\mciteSetBstMidEndSepPunct{\mcitedefaultmidpunct}
{\mcitedefaultendpunct}{\mcitedefaultseppunct}\relax
\EndOfBibitem
\bibitem[Smith \latin{et~al.}(2017)Smith, Isayev, and Roitberg]{smith2017ani}
Smith,~J.~S.; Isayev,~O.; Roitberg,~A.~E. ANI-1: an extensible neural network
  potential with DFT accuracy at force field computational cost. \emph{Chemical
  science} \textbf{2017}, \emph{8}, 3192--3203\relax
\mciteBstWouldAddEndPuncttrue
\mciteSetBstMidEndSepPunct{\mcitedefaultmidpunct}
{\mcitedefaultendpunct}{\mcitedefaultseppunct}\relax
\EndOfBibitem
\bibitem[Klicpera \latin{et~al.}(2020)Klicpera, Giri, Margraf, and
  G{\"u}nnemann]{klicpera2020fast}
Klicpera,~J.; Giri,~S.; Margraf,~J.~T.; G{\"u}nnemann,~S. Fast and
  Uncertainty-Aware Directional Message Passing for Non-Equilibrium Molecules.
  \emph{arXiv preprint arXiv:2011.14115} \textbf{2020}, \relax
\mciteBstWouldAddEndPunctfalse
\mciteSetBstMidEndSepPunct{\mcitedefaultmidpunct}
{}{\mcitedefaultseppunct}\relax
\EndOfBibitem
\bibitem[Ramakrishnan \latin{et~al.}(2014)Ramakrishnan, Dral, Rupp, and
  Von~Lilienfeld]{ramakrishnan2014quantum}
Ramakrishnan,~R.; Dral,~P.~O.; Rupp,~M.; Von~Lilienfeld,~O.~A. Quantum
  chemistry structures and properties of 134 kilo molecules. \emph{Scientific
  data} \textbf{2014}, \emph{1}, 1--7\relax
\mciteBstWouldAddEndPuncttrue
\mciteSetBstMidEndSepPunct{\mcitedefaultmidpunct}
{\mcitedefaultendpunct}{\mcitedefaultseppunct}\relax
\EndOfBibitem
\bibitem[Chen \latin{et~al.}(2019)Chen, Chen, Hsieh, Lee, Liao, Liao, Liu, Qiu,
  Sun, Tang, \latin{et~al.} others]{chen2019alchemy}
Chen,~G.; Chen,~P.; Hsieh,~C.-Y.; Lee,~C.-K.; Liao,~B.; Liao,~R.; Liu,~W.;
  Qiu,~J.; Sun,~Q.; Tang,~J., \latin{et~al.}  Alchemy: A quantum chemistry
  dataset for benchmarking ai models. \emph{arXiv preprint arXiv:1906.09427}
  \textbf{2019}, \relax
\mciteBstWouldAddEndPunctfalse
\mciteSetBstMidEndSepPunct{\mcitedefaultmidpunct}
{}{\mcitedefaultseppunct}\relax
\EndOfBibitem
\bibitem[Andersen \latin{et~al.}(2019)Andersen, Levchenko, Scheffler, and
  Reuter]{andersen2019beyond}
Andersen,~M.; Levchenko,~S.~V.; Scheffler,~M.; Reuter,~K. Beyond scaling
  relations for the description of catalytic materials. \emph{ACS Catalysis}
  \textbf{2019}, \emph{9}, 2752--2759\relax
\mciteBstWouldAddEndPuncttrue
\mciteSetBstMidEndSepPunct{\mcitedefaultmidpunct}
{\mcitedefaultendpunct}{\mcitedefaultseppunct}\relax
\EndOfBibitem
\bibitem[Abild-Pedersen \latin{et~al.}(2007)Abild-Pedersen, Greeley, Studt,
  Rossmeisl, Munter, Moses, Skulason, Bligaard, and
  N{\o}rskov]{abild2007scaling}
Abild-Pedersen,~F.; Greeley,~J.; Studt,~F.; Rossmeisl,~J.; Munter,~T.~R.;
  Moses,~P.~G.; Skulason,~E.; Bligaard,~T.; N{\o}rskov,~J.~K. Scaling
  properties of adsorption energies for hydrogen-containing molecules on
  transition-metal surfaces. \emph{Physical review letters} \textbf{2007},
  \emph{99}, 016105\relax
\mciteBstWouldAddEndPuncttrue
\mciteSetBstMidEndSepPunct{\mcitedefaultmidpunct}
{\mcitedefaultendpunct}{\mcitedefaultseppunct}\relax
\EndOfBibitem
\bibitem[Ma and Xin(2017)Ma, and Xin]{ma2017orbitalwise}
Ma,~X.; Xin,~H. Orbitalwise coordination number for predicting adsorption
  properties of metal nanocatalysts. \emph{Physical review letters}
  \textbf{2017}, \emph{118}, 036101\relax
\mciteBstWouldAddEndPuncttrue
\mciteSetBstMidEndSepPunct{\mcitedefaultmidpunct}
{\mcitedefaultendpunct}{\mcitedefaultseppunct}\relax
\EndOfBibitem
\bibitem[Noh \latin{et~al.}(2018)Noh, Back, Kim, and Jung]{noh2018active}
Noh,~J.; Back,~S.; Kim,~J.; Jung,~Y. Active learning with non-ab initio input
  features toward efficient \ce{CO2} reduction catalysts. \emph{Chemical
  science} \textbf{2018}, \emph{9}, 5152--5159\relax
\mciteBstWouldAddEndPuncttrue
\mciteSetBstMidEndSepPunct{\mcitedefaultmidpunct}
{\mcitedefaultendpunct}{\mcitedefaultseppunct}\relax
\EndOfBibitem
\bibitem[Behler and Parrinello(2007)Behler, and
  Parrinello]{behler2007generalized}
Behler,~J.; Parrinello,~M. Generalized neural-network representation of
  high-dimensional potential-energy surfaces. \emph{Physical review letters}
  \textbf{2007}, \emph{98}, 146401\relax
\mciteBstWouldAddEndPuncttrue
\mciteSetBstMidEndSepPunct{\mcitedefaultmidpunct}
{\mcitedefaultendpunct}{\mcitedefaultseppunct}\relax
\EndOfBibitem
\bibitem[Lorenz \latin{et~al.}(2004)Lorenz, Gro{\ss}, and
  Scheffler]{lorenz2004representing}
Lorenz,~S.; Gro{\ss},~A.; Scheffler,~M. Representing high-dimensional
  potential-energy surfaces for reactions at surfaces by neural networks.
  \emph{Chemical Physics Letters} \textbf{2004}, \emph{395}, 210--215\relax
\mciteBstWouldAddEndPuncttrue
\mciteSetBstMidEndSepPunct{\mcitedefaultmidpunct}
{\mcitedefaultendpunct}{\mcitedefaultseppunct}\relax
\EndOfBibitem
\bibitem[Chen \latin{et~al.}(2013)Chen, Xu, Xu, and Zhang]{chen2013global}
Chen,~J.; Xu,~X.; Xu,~X.; Zhang,~D.~H. A global potential energy surface for
  the \ce{H2 + OH -> H2O + H} reaction using neural networks. \emph{The Journal
  of Chemical Physics} \textbf{2013}, \emph{138}, 154301\relax
\mciteBstWouldAddEndPuncttrue
\mciteSetBstMidEndSepPunct{\mcitedefaultmidpunct}
{\mcitedefaultendpunct}{\mcitedefaultseppunct}\relax
\EndOfBibitem
\bibitem[Bart{\'o}k and Cs{\'a}nyi(2015)Bart{\'o}k, and
  Cs{\'a}nyi]{bartok2015g}
Bart{\'o}k,~A.~P.; Cs{\'a}nyi,~G. G aussian approximation potentials: A brief
  tutorial introduction. \emph{International Journal of Quantum Chemistry}
  \textbf{2015}, \emph{115}, 1051--1057\relax
\mciteBstWouldAddEndPuncttrue
\mciteSetBstMidEndSepPunct{\mcitedefaultmidpunct}
{\mcitedefaultendpunct}{\mcitedefaultseppunct}\relax
\EndOfBibitem
\bibitem[Bart{\'o}k \latin{et~al.}(2010)Bart{\'o}k, Payne, Kondor, and
  Cs{\'a}nyi]{bartok2010gaussian}
Bart{\'o}k,~A.~P.; Payne,~M.~C.; Kondor,~R.; Cs{\'a}nyi,~G. Gaussian
  approximation potentials: The accuracy of quantum mechanics, without the
  electrons. \emph{Physical review letters} \textbf{2010}, \emph{104},
  136403\relax
\mciteBstWouldAddEndPuncttrue
\mciteSetBstMidEndSepPunct{\mcitedefaultmidpunct}
{\mcitedefaultendpunct}{\mcitedefaultseppunct}\relax
\EndOfBibitem
\bibitem[Sch{\"u}tt \latin{et~al.}(2017)Sch{\"u}tt, Kindermans, Felix, Chmiela,
  Tkatchenko, and M{\"u}ller]{schutt2017schnet}
Sch{\"u}tt,~K.; Kindermans,~P.-J.; Felix,~H. E.~S.; Chmiela,~S.;
  Tkatchenko,~A.; M{\"u}ller,~K.-R. Schnet: A continuous-filter convolutional
  neural network for modeling quantum interactions. NeurIPS. 2017; pp
  991--1001\relax
\mciteBstWouldAddEndPuncttrue
\mciteSetBstMidEndSepPunct{\mcitedefaultmidpunct}
{\mcitedefaultendpunct}{\mcitedefaultseppunct}\relax
\EndOfBibitem
\bibitem[Klicpera \latin{et~al.}(2020)Klicpera, Gro{\ss}, and
  G{\"u}nnemann]{klicpera2020directional}
Klicpera,~J.; Gro{\ss},~J.; G{\"u}nnemann,~S. Directional message passing for
  molecular graphs. ICLR. 2020\relax
\mciteBstWouldAddEndPuncttrue
\mciteSetBstMidEndSepPunct{\mcitedefaultmidpunct}
{\mcitedefaultendpunct}{\mcitedefaultseppunct}\relax
\EndOfBibitem
\bibitem[Klicpera \latin{et~al.}(2021)Klicpera, Becker, and
  G{\"u}nnemann]{klicpera2021gemnet}
Klicpera,~J.; Becker,~F.; G{\"u}nnemann,~S. GemNet: Universal Directional Graph
  Neural Networks for Molecules. \emph{arXiv preprint arXiv:2106.08903}
  \textbf{2021}, \relax
\mciteBstWouldAddEndPunctfalse
\mciteSetBstMidEndSepPunct{\mcitedefaultmidpunct}
{}{\mcitedefaultseppunct}\relax
\EndOfBibitem
\bibitem[Batzner \latin{et~al.}(2021)Batzner, Musaelian, Sun, Geiger, Mailoa,
  Kornbluth, Molinari, Smidt, and Kozinsky]{batzner2021se}
Batzner,~S.; Musaelian,~A.; Sun,~L.; Geiger,~M.; Mailoa,~J.~P.; Kornbluth,~M.;
  Molinari,~N.; Smidt,~T.~E.; Kozinsky,~B. Se (3)-equivariant graph neural
  networks for data-efficient and accurate interatomic potentials. \emph{arXiv
  preprint arXiv:2101.03164} \textbf{2021}, \relax
\mciteBstWouldAddEndPunctfalse
\mciteSetBstMidEndSepPunct{\mcitedefaultmidpunct}
{}{\mcitedefaultseppunct}\relax
\EndOfBibitem
\bibitem[Sch{\"u}tt \latin{et~al.}(2021)Sch{\"u}tt, Unke, and
  Gastegger]{schutt2021equivariant}
Sch{\"u}tt,~K.~T.; Unke,~O.~T.; Gastegger,~M. Equivariant message passing for
  the prediction of tensorial properties and molecular spectra. \emph{arXiv
  preprint arXiv:2102.03150} \textbf{2021}, \relax
\mciteBstWouldAddEndPunctfalse
\mciteSetBstMidEndSepPunct{\mcitedefaultmidpunct}
{}{\mcitedefaultseppunct}\relax
\EndOfBibitem
\bibitem[Liu \latin{et~al.}(2021)Liu, Wang, Liu, Zhang, Oztekin, and
  Ji]{liu2021spherical}
Liu,~Y.; Wang,~L.; Liu,~M.; Zhang,~X.; Oztekin,~B.; Ji,~S. Spherical message
  passing for 3d graph networks. \emph{arXiv preprint arXiv:2102.05013}
  \textbf{2021}, \relax
\mciteBstWouldAddEndPunctfalse
\mciteSetBstMidEndSepPunct{\mcitedefaultmidpunct}
{}{\mcitedefaultseppunct}\relax
\EndOfBibitem
\bibitem[Lei and Medford(2021)Lei, and Medford]{lei2021universal}
Lei,~X.; Medford,~A.~J. A universal framework for featurization of atomistic
  systems. \emph{arXiv preprint arXiv:2102.02390} \textbf{2021}, \relax
\mciteBstWouldAddEndPunctfalse
\mciteSetBstMidEndSepPunct{\mcitedefaultmidpunct}
{}{\mcitedefaultseppunct}\relax
\EndOfBibitem
\bibitem[Shuaibi \latin{et~al.}(2021)Shuaibi, Kolluru, Das, Grover, Sriram,
  Ulissi, and Zitnick]{shuaibi2021rotation}
Shuaibi,~M.; Kolluru,~A.; Das,~A.; Grover,~A.; Sriram,~A.; Ulissi,~Z.;
  Zitnick,~C.~L. Rotation Invariant Graph Neural Networks using Spin
  Convolutions. \emph{arXiv preprint arXiv:2106.09575} \textbf{2021}, \relax
\mciteBstWouldAddEndPunctfalse
\mciteSetBstMidEndSepPunct{\mcitedefaultmidpunct}
{}{\mcitedefaultseppunct}\relax
\EndOfBibitem
\bibitem[Sriram \latin{et~al.}(2021)Sriram, Das, Wood, and
  Zitnick]{sriram2021towards}
Sriram,~A.; Das,~A.; Wood,~B.~M.; Zitnick,~C.~L. Towards Training Billion
  Parameter Graph Neural Networks for Atomic Simulations. International
  Conference on Learning Representations. 2021\relax
\mciteBstWouldAddEndPuncttrue
\mciteSetBstMidEndSepPunct{\mcitedefaultmidpunct}
{\mcitedefaultendpunct}{\mcitedefaultseppunct}\relax
\EndOfBibitem
\bibitem[Ying \latin{et~al.}(2021)Ying, Cai, Luo, Zheng, Ke, He, Shen, and
  Liu]{ying2021transformers}
Ying,~C.; Cai,~T.; Luo,~S.; Zheng,~S.; Ke,~G.; He,~D.; Shen,~Y.; Liu,~T.-Y. Do
  Transformers Really Perform Badly for Graph Representation? \emph{Advances in
  Neural Information Processing Systems} \textbf{2021}, \emph{34}\relax
\mciteBstWouldAddEndPuncttrue
\mciteSetBstMidEndSepPunct{\mcitedefaultmidpunct}
{\mcitedefaultendpunct}{\mcitedefaultseppunct}\relax
\EndOfBibitem
\bibitem[Godwin \latin{et~al.}(2021)Godwin, Schaarschmidt, Gaunt,
  Sanchez-Gonzalez, Rubanova, Veli{\v{c}}kovi{\'c}, Kirkpatrick, and
  Battaglia]{godwin2021simple}
Godwin,~J.; Schaarschmidt,~M.; Gaunt,~A.~L.; Sanchez-Gonzalez,~A.;
  Rubanova,~Y.; Veli{\v{c}}kovi{\'c},~P.; Kirkpatrick,~J.; Battaglia,~P. Simple
  GNN Regularisation for 3D Molecular Property Prediction and Beyond.
  International Conference on Learning Representations. 2021\relax
\mciteBstWouldAddEndPuncttrue
\mciteSetBstMidEndSepPunct{\mcitedefaultmidpunct}
{\mcitedefaultendpunct}{\mcitedefaultseppunct}\relax
\EndOfBibitem
\bibitem[ocp(2021)]{ocpchallenge}
Open Catalyst Project Challenge.
  \url{https://opencatalystproject.org/challenge.html}, 2021\relax
\mciteBstWouldAddEndPuncttrue
\mciteSetBstMidEndSepPunct{\mcitedefaultmidpunct}
{\mcitedefaultendpunct}{\mcitedefaultseppunct}\relax
\EndOfBibitem
\bibitem[Gasteiger \latin{et~al.}(2022)Gasteiger, Shuaibi, Sriram, Günnemann,
  Ulissi, Zitnick, and Das]{gemnetoc}
Gasteiger,~J.; Shuaibi,~M.; Sriram,~A.; Günnemann,~S.; Ulissi,~Z.;
  Zitnick,~C.~L.; Das,~A. How Do Graph Networks Generalize to Large and Diverse
  Molecular Systems? 2022; \url{https://arxiv.org/abs/2204.02782}\relax
\mciteBstWouldAddEndPuncttrue
\mciteSetBstMidEndSepPunct{\mcitedefaultmidpunct}
{\mcitedefaultendpunct}{\mcitedefaultseppunct}\relax
\EndOfBibitem
\bibitem[Kolluru \latin{et~al.}(2022)Kolluru, Shoghi, Shuaibi, Goyal, Das,
  Zitnick, and Ulissi]{kolluru2022transfer}
Kolluru,~A.; Shoghi,~N.; Shuaibi,~M.; Goyal,~S.; Das,~A.; Zitnick,~L.;
  Ulissi,~Z.~W. Transfer Learning using Attentions across Atomic Systems with
  Graph Neural Networks (TAAG). \emph{The Journal of Chemical Physics}
  \textbf{2022}, \relax
\mciteBstWouldAddEndPunctfalse
\mciteSetBstMidEndSepPunct{\mcitedefaultmidpunct}
{}{\mcitedefaultseppunct}\relax
\EndOfBibitem
\bibitem[Chen and Ong(2021)Chen, and Ong]{chen2021atomsets}
Chen,~C.; Ong,~S.~P. AtomSets as a hierarchical transfer learning framework for
  small and large materials datasets. \emph{npj Computational Materials}
  \textbf{2021}, \emph{7}, 1--9\relax
\mciteBstWouldAddEndPuncttrue
\mciteSetBstMidEndSepPunct{\mcitedefaultmidpunct}
{\mcitedefaultendpunct}{\mcitedefaultseppunct}\relax
\EndOfBibitem
\bibitem[Back \latin{et~al.}(2019)Back, Yoon, Tian, Zhong, Tran, and
  Ulissi]{back2019convolutional}
Back,~S.; Yoon,~J.; Tian,~N.; Zhong,~W.; Tran,~K.; Ulissi,~Z.~W. Convolutional
  neural network of atomic surface structures to predict binding energies for
  high-throughput screening of catalysts. \emph{The journal of physical
  chemistry letters} \textbf{2019}, \emph{10}, 4401--4408\relax
\mciteBstWouldAddEndPuncttrue
\mciteSetBstMidEndSepPunct{\mcitedefaultmidpunct}
{\mcitedefaultendpunct}{\mcitedefaultseppunct}\relax
\EndOfBibitem
\bibitem[Yoon and Ulissi(2020)Yoon, and Ulissi]{yoon2020differentiable}
Yoon,~J.; Ulissi,~Z.~W. Differentiable optimization for the prediction of
  ground state structures (DOGSS). \emph{Physical Review Letters}
  \textbf{2020}, \emph{125}, 173001\relax
\mciteBstWouldAddEndPuncttrue
\mciteSetBstMidEndSepPunct{\mcitedefaultmidpunct}
{\mcitedefaultendpunct}{\mcitedefaultseppunct}\relax
\EndOfBibitem
\bibitem[Chen \latin{et~al.}(2020)Chen, Zhang, Chen, Yao, and
  Zhou]{chen2020machine}
Chen,~A.; Zhang,~X.; Chen,~L.; Yao,~S.; Zhou,~Z. A machine learning model on
  simple features for \ce{CO2} reduction electrocatalysts. \emph{The Journal of
  Physical Chemistry C} \textbf{2020}, \emph{124}, 22471--22478\relax
\mciteBstWouldAddEndPuncttrue
\mciteSetBstMidEndSepPunct{\mcitedefaultmidpunct}
{\mcitedefaultendpunct}{\mcitedefaultseppunct}\relax
\EndOfBibitem
\bibitem[dis(2021)]{disc_post}
IS2RE Leaderboard Concerns.
  \url{https://discuss.opencatalystproject.org/t/is2re-leaderboard-concerns/66},
  2021\relax
\mciteBstWouldAddEndPuncttrue
\mciteSetBstMidEndSepPunct{\mcitedefaultmidpunct}
{\mcitedefaultendpunct}{\mcitedefaultseppunct}\relax
\EndOfBibitem
\bibitem[Li \latin{et~al.}(2017)Li, Wang, Chin, Achenie, and Xin]{li2017high}
Li,~Z.; Wang,~S.; Chin,~W.~S.; Achenie,~L.~E.; Xin,~H. High-throughput
  screening of bimetallic catalysts enabled by machine learning. \emph{Journal
  of Materials Chemistry A} \textbf{2017}, \emph{5}, 24131--24138\relax
\mciteBstWouldAddEndPuncttrue
\mciteSetBstMidEndSepPunct{\mcitedefaultmidpunct}
{\mcitedefaultendpunct}{\mcitedefaultseppunct}\relax
\EndOfBibitem
\bibitem[Mamun \latin{et~al.}(2019)Mamun, Winther, Boes, and
  Bligaard]{mamun2019high}
Mamun,~O.; Winther,~K.~T.; Boes,~J.~R.; Bligaard,~T. High-throughput
  calculations of catalytic properties of bimetallic alloy surfaces.
  \emph{Scientific data} \textbf{2019}, \emph{6}, 1--9\relax
\mciteBstWouldAddEndPuncttrue
\mciteSetBstMidEndSepPunct{\mcitedefaultmidpunct}
{\mcitedefaultendpunct}{\mcitedefaultseppunct}\relax
\EndOfBibitem
\bibitem[Peterson(2016)]{peterson2016acceleration}
Peterson,~A.~A. Acceleration of saddle-point searches with machine learning.
  \emph{The Journal of chemical physics} \textbf{2016}, \emph{145},
  074106\relax
\mciteBstWouldAddEndPuncttrue
\mciteSetBstMidEndSepPunct{\mcitedefaultmidpunct}
{\mcitedefaultendpunct}{\mcitedefaultseppunct}\relax
\EndOfBibitem
\bibitem[Christensen and von Lilienfeld(2020)Christensen, and von
  Lilienfeld]{christensen2020role}
Christensen,~A.~S.; von Lilienfeld,~O.~A. On the role of gradients for machine
  learning of molecular energies and forces. \emph{Machine Learning: Science
  and Technology} \textbf{2020}, \emph{1}, 045018\relax
\mciteBstWouldAddEndPuncttrue
\mciteSetBstMidEndSepPunct{\mcitedefaultmidpunct}
{\mcitedefaultendpunct}{\mcitedefaultseppunct}\relax
\EndOfBibitem
\bibitem[Hu \latin{et~al.}(2021)Hu, Shuaibi, Das, Goyal, Sriram, Leskovec,
  Parikh, and Zitnick]{hu2021forcenet}
Hu,~W.; Shuaibi,~M.; Das,~A.; Goyal,~S.; Sriram,~A.; Leskovec,~J.; Parikh,~D.;
  Zitnick,~C.~L. Forcenet: A graph neural network for large-scale quantum
  calculations. \emph{arXiv preprint arXiv:2103.01436} \textbf{2021}, \relax
\mciteBstWouldAddEndPunctfalse
\mciteSetBstMidEndSepPunct{\mcitedefaultmidpunct}
{}{\mcitedefaultseppunct}\relax
\EndOfBibitem
\bibitem[del R{\'\i}o \latin{et~al.}(2019)del R{\'\i}o, Mortensen, and
  Jacobsen]{del2019local}
del R{\'\i}o,~E.~G.; Mortensen,~J.~J.; Jacobsen,~K.~W. Local Bayesian optimizer
  for atomic structures. \emph{Physical Review B} \textbf{2019}, \emph{100},
  104103\relax
\mciteBstWouldAddEndPuncttrue
\mciteSetBstMidEndSepPunct{\mcitedefaultmidpunct}
{\mcitedefaultendpunct}{\mcitedefaultseppunct}\relax
\EndOfBibitem
\bibitem[Wang \latin{et~al.}(2018)Wang, Qiu, Song, Gu, Zheng, Zhao, Zhao, Liu,
  and Zhang]{wang2018adsorption}
Wang,~Y.; Qiu,~W.; Song,~E.; Gu,~F.; Zheng,~Z.; Zhao,~X.; Zhao,~Y.; Liu,~J.;
  Zhang,~W. Adsorption-energy-based activity descriptors for electrocatalysts
  in energy storage applications. \emph{National Science Review} \textbf{2018},
  \emph{5}, 327--341\relax
\mciteBstWouldAddEndPuncttrue
\mciteSetBstMidEndSepPunct{\mcitedefaultmidpunct}
{\mcitedefaultendpunct}{\mcitedefaultseppunct}\relax
\EndOfBibitem
\bibitem[Garc{\'\i}a-Muelas and L{\'o}pez(2019)Garc{\'\i}a-Muelas, and
  L{\'o}pez]{garcia2019statistical}
Garc{\'\i}a-Muelas,~R.; L{\'o}pez,~N. Statistical learning goes beyond the
  d-band model providing the thermochemistry of adsorbates on transition
  metals. \emph{Nature communications} \textbf{2019}, \emph{10}, 1--7\relax
\mciteBstWouldAddEndPuncttrue
\mciteSetBstMidEndSepPunct{\mcitedefaultmidpunct}
{\mcitedefaultendpunct}{\mcitedefaultseppunct}\relax
\EndOfBibitem
\bibitem[Jumper \latin{et~al.}(2021)Jumper, Evans, Pritzel, Green, Figurnov,
  Ronneberger, Tunyasuvunakool, Bates, {\v{Z}}{\'\i}dek, Potapenko,
  \latin{et~al.} others]{jumper2021highly}
Jumper,~J.; Evans,~R.; Pritzel,~A.; Green,~T.; Figurnov,~M.; Ronneberger,~O.;
  Tunyasuvunakool,~K.; Bates,~R.; {\v{Z}}{\'\i}dek,~A.; Potapenko,~A.,
  \latin{et~al.}  Highly accurate protein structure prediction with AlphaFold.
  \emph{Nature} \textbf{2021}, \emph{596}, 583--589\relax
\mciteBstWouldAddEndPuncttrue
\mciteSetBstMidEndSepPunct{\mcitedefaultmidpunct}
{\mcitedefaultendpunct}{\mcitedefaultseppunct}\relax
\EndOfBibitem
\bibitem[Zemla(2003)]{zemla2003lga}
Zemla,~A. LGA: a method for finding 3D similarities in protein structures.
  \emph{Nucleic acids research} \textbf{2003}, \emph{31}, 3370--3374\relax
\mciteBstWouldAddEndPuncttrue
\mciteSetBstMidEndSepPunct{\mcitedefaultmidpunct}
{\mcitedefaultendpunct}{\mcitedefaultseppunct}\relax
\EndOfBibitem
\bibitem[Tran \latin{et~al.}(2020)Tran, Neiswanger, Yoon, Zhang, Xing, and
  Ulissi]{tran2020methods}
Tran,~K.; Neiswanger,~W.; Yoon,~J.; Zhang,~Q.; Xing,~E.; Ulissi,~Z.~W. Methods
  for comparing uncertainty quantifications for material property predictions.
  \emph{Machine Learning: Science and Technology} \textbf{2020}, \emph{1},
  025006\relax
\mciteBstWouldAddEndPuncttrue
\mciteSetBstMidEndSepPunct{\mcitedefaultmidpunct}
{\mcitedefaultendpunct}{\mcitedefaultseppunct}\relax
\EndOfBibitem
\bibitem[Busk \latin{et~al.}(2021)Busk, J{\o}rgensen, Bhowmik, Schmidt,
  Winther, and Vegge]{busk2021calibrated}
Busk,~J.; J{\o}rgensen,~P.~B.; Bhowmik,~A.; Schmidt,~M.~N.; Winther,~O.;
  Vegge,~T. Calibrated uncertainty for molecular property prediction using
  ensembles of message passing neural networks. \emph{Machine Learning: Science
  and Technology} \textbf{2021}, \emph{3}, 015012\relax
\mciteBstWouldAddEndPuncttrue
\mciteSetBstMidEndSepPunct{\mcitedefaultmidpunct}
{\mcitedefaultendpunct}{\mcitedefaultseppunct}\relax
\EndOfBibitem
\bibitem[Shuaibi \latin{et~al.}(2020)Shuaibi, Sivakumar, Chen, and
  Ulissi]{shuaibi2020enabling}
Shuaibi,~M.; Sivakumar,~S.; Chen,~R.~Q.; Ulissi,~Z.~W. Enabling robust offline
  active learning for machine learning potentials using simple physics-based
  priors. \emph{Machine Learning: Science and Technology} \textbf{2020},
  \emph{2}, 025007\relax
\mciteBstWouldAddEndPuncttrue
\mciteSetBstMidEndSepPunct{\mcitedefaultmidpunct}
{\mcitedefaultendpunct}{\mcitedefaultseppunct}\relax
\EndOfBibitem
\bibitem[Vandermause \latin{et~al.}(2020)Vandermause, Torrisi, Batzner, Xie,
  Sun, Kolpak, and Kozinsky]{vandermause2020fly}
Vandermause,~J.; Torrisi,~S.~B.; Batzner,~S.; Xie,~Y.; Sun,~L.; Kolpak,~A.~M.;
  Kozinsky,~B. On-the-fly active learning of interpretable Bayesian force
  fields for atomistic rare events. \emph{npj Computational Materials}
  \textbf{2020}, \emph{6}, 1--11\relax
\mciteBstWouldAddEndPuncttrue
\mciteSetBstMidEndSepPunct{\mcitedefaultmidpunct}
{\mcitedefaultendpunct}{\mcitedefaultseppunct}\relax
\EndOfBibitem
\bibitem[Jinnouchi \latin{et~al.}(2019)Jinnouchi, Lahnsteiner, Karsai, Kresse,
  and Bokdam]{jinnouchi2019phase}
Jinnouchi,~R.; Lahnsteiner,~J.; Karsai,~F.; Kresse,~G.; Bokdam,~M. Phase
  transitions of hybrid perovskites simulated by machine-learning force fields
  trained on the fly with Bayesian inference. \emph{Physical review letters}
  \textbf{2019}, \emph{122}, 225701\relax
\mciteBstWouldAddEndPuncttrue
\mciteSetBstMidEndSepPunct{\mcitedefaultmidpunct}
{\mcitedefaultendpunct}{\mcitedefaultseppunct}\relax
\EndOfBibitem
\bibitem[Torres \latin{et~al.}(2019)Torres, Jennings, Hansen, Boes, and
  Bligaard]{torres2019low}
Torres,~J. A.~G.; Jennings,~P.~C.; Hansen,~M.~H.; Boes,~J.~R.; Bligaard,~T.
  Low-scaling algorithm for nudged elastic band calculations using a surrogate
  machine learning model. \emph{Physical review letters} \textbf{2019},
  \emph{122}, 156001\relax
\mciteBstWouldAddEndPuncttrue
\mciteSetBstMidEndSepPunct{\mcitedefaultmidpunct}
{\mcitedefaultendpunct}{\mcitedefaultseppunct}\relax
\EndOfBibitem
\bibitem[Scalia \latin{et~al.}(2020)Scalia, Grambow, Pernici, Li, and
  Green]{scalia2020evaluating}
Scalia,~G.; Grambow,~C.~A.; Pernici,~B.; Li,~Y.-P.; Green,~W.~H. Evaluating
  scalable uncertainty estimation methods for deep learning-based molecular
  property prediction. \emph{Journal of chemical information and modeling}
  \textbf{2020}, \emph{60}, 2697--2717\relax
\mciteBstWouldAddEndPuncttrue
\mciteSetBstMidEndSepPunct{\mcitedefaultmidpunct}
{\mcitedefaultendpunct}{\mcitedefaultseppunct}\relax
\EndOfBibitem
\bibitem[Smith \latin{et~al.}(2020)Smith, Zubatyuk, Nebgen, Lubbers, Barros,
  Roitberg, Isayev, and Tretiak]{smith2020ani}
Smith,~J.~S.; Zubatyuk,~R.; Nebgen,~B.; Lubbers,~N.; Barros,~K.;
  Roitberg,~A.~E.; Isayev,~O.; Tretiak,~S. The ANI-1ccx and ANI-1x data sets,
  coupled-cluster and density functional theory properties for molecules.
  \emph{Scientific data} \textbf{2020}, \emph{7}, 1--10\relax
\mciteBstWouldAddEndPuncttrue
\mciteSetBstMidEndSepPunct{\mcitedefaultmidpunct}
{\mcitedefaultendpunct}{\mcitedefaultseppunct}\relax
\EndOfBibitem
\bibitem[Musaelian \latin{et~al.}(2022)Musaelian, Batzner, Johansson, Sun,
  Owen, Kornbluth, and Kozinsky]{musaelian2022learning}
Musaelian,~A.; Batzner,~S.; Johansson,~A.; Sun,~L.; Owen,~C.~J.; Kornbluth,~M.;
  Kozinsky,~B. Learning Local Equivariant Representations for Large-Scale
  Atomistic Dynamics. \emph{arXiv preprint arXiv:2204.05249} \textbf{2022},
  \relax
\mciteBstWouldAddEndPunctfalse
\mciteSetBstMidEndSepPunct{\mcitedefaultmidpunct}
{}{\mcitedefaultseppunct}\relax
\EndOfBibitem
\bibitem[Frankle and Carbin(2018)Frankle, and Carbin]{frankle2018lottery}
Frankle,~J.; Carbin,~M. The lottery ticket hypothesis: Finding sparse,
  trainable neural networks. \emph{arXiv preprint arXiv:1803.03635}
  \textbf{2018}, \relax
\mciteBstWouldAddEndPunctfalse
\mciteSetBstMidEndSepPunct{\mcitedefaultmidpunct}
{}{\mcitedefaultseppunct}\relax
\EndOfBibitem
\bibitem[Kirklin \latin{et~al.}(2015)Kirklin, Saal, Meredig, Thompson, Doak,
  Aykol, R{\"u}hl, and Wolverton]{kirklin2015open}
Kirklin,~S.; Saal,~J.~E.; Meredig,~B.; Thompson,~A.; Doak,~J.~W.; Aykol,~M.;
  R{\"u}hl,~S.; Wolverton,~C. The Open Quantum Materials Database (OQMD):
  assessing the accuracy of DFT formation energies. \emph{npj Computational
  Materials} \textbf{2015}, \emph{1}, 1--15\relax
\mciteBstWouldAddEndPuncttrue
\mciteSetBstMidEndSepPunct{\mcitedefaultmidpunct}
{\mcitedefaultendpunct}{\mcitedefaultseppunct}\relax
\EndOfBibitem
\bibitem[Qiao \latin{et~al.}(2021)Qiao, Christensen, Welborn, Manby,
  Anandkumar, and Miller~III]{qiao2021unite}
Qiao,~Z.; Christensen,~A.~S.; Welborn,~M.; Manby,~F.~R.; Anandkumar,~A.;
  Miller~III,~T.~F. Unite: Unitary n-body tensor equivariant network with
  applications to quantum chemistry. \emph{arXiv preprint arXiv:2105.14655}
  \textbf{2021}, \relax
\mciteBstWouldAddEndPunctfalse
\mciteSetBstMidEndSepPunct{\mcitedefaultmidpunct}
{}{\mcitedefaultseppunct}\relax
\EndOfBibitem
\bibitem[Qiao \latin{et~al.}(2020)Qiao, Welborn, Anandkumar, Manby, and
  Miller~III]{qiao2020orbnet}
Qiao,~Z.; Welborn,~M.; Anandkumar,~A.; Manby,~F.~R.; Miller~III,~T.~F. OrbNet:
  Deep learning for quantum chemistry using symmetry-adapted atomic-orbital
  features. \emph{The Journal of Chemical Physics} \textbf{2020}, \emph{153},
  124111\relax
\mciteBstWouldAddEndPuncttrue
\mciteSetBstMidEndSepPunct{\mcitedefaultmidpunct}
{\mcitedefaultendpunct}{\mcitedefaultseppunct}\relax
\EndOfBibitem
\bibitem[Bannwarth \latin{et~al.}(2019)Bannwarth, Ehlert, and
  Grimme]{bannwarth2019gfn2}
Bannwarth,~C.; Ehlert,~S.; Grimme,~S. GFN2-xTB—An accurate and broadly
  parametrized self-consistent tight-binding quantum chemical method with
  multipole electrostatics and density-dependent dispersion contributions.
  \emph{Journal of chemical theory and computation} \textbf{2019}, \emph{15},
  1652--1671\relax
\mciteBstWouldAddEndPuncttrue
\mciteSetBstMidEndSepPunct{\mcitedefaultmidpunct}
{\mcitedefaultendpunct}{\mcitedefaultseppunct}\relax
\EndOfBibitem
\end{mcitethebibliography}

\end{document}